\documentclass[nature, aps, pra, 10pt]{revtex4-1}

\usepackage[svgnames]{xcolor} % Specify colors by their 'svgnames', for a full list of all colors available see here: http://www.latextemplates.com/svgnames-colors
\usepackage{graphicx,epsfig}
\usepackage{times}
\usepackage{amssymb,amsmath,multirow,rotate,color}
\usepackage{xcolor}
\usepackage{physics}
\usepackage[normalem]{ulem}
\usepackage{hyperref}
\usepackage{rotating}
\usepackage{lineno}
\linenumbers
\usepackage{titlesec}
\usepackage{setspace} %%%7/31/24 4:54pm REMOVE AFTR EDITING

\newcommand{\rev}[1]{\textcolor{black}{#1}}

\begin{document}

\nolinenumbers

%\title{Small subset of high quality connections preserves disease spreading}
%\title{Relevance of shortest paths and edge redundancy for disease spreading}

\title{Quantifying edge relevance for epidemic spreading via the semi-metric topology of complex networks}

\author{David Soriano-Pa\~nos}
\email{sorianopanos@gmail.com}
\affiliation{Departament d'Enginyeria Inform\`atica i Matem\`atiques, Universitat Rovira i Virgili, 43007 Tarragona (Spain).}
\affiliation{GOTHAM lab, Institute for Biocomputation and Physics of Complex Systems, University of Zaragoza,
50018 Zaragoza, Spain.}

\author{Felipe Xavier Costa}
%\affiliation{Instituto Gulbenkian de Ci\^encia, 2780-156 Oeiras, Portugal.}
\affiliation{Universidade Católica Portuguesa, Católica Medical School, Católica Biomedical Research Centre, Portugal}
\affiliation{School of Systems Science and Industrial Engineering, Binghamton University (State University of New York), Binghamton, NY 13902, USA}
%\affiliation{Department of Physics, State University of New York at Albany, Albany, NY 12222, USA.}

\author{Luis M. Rocha}
\email{rocha@binghamton.edu}
\affiliation{School of Systems Science and Industrial Engineering, Binghamton University (State University of New York), Binghamton, NY 13902, USA.}
\affiliation{Universidade Católica Portuguesa, Católica Medical School, Católica Biomedical Research Centre, Portugal}

\date{\today}

\begin{abstract}

%
%Sparsification is an essential tool for extracting a reduced core of associations that are sufficient to maintain both the dynamics and topology of networks. 
%
\rev{Sparsification aims at extracting a reduced core of associations that best preserves both the dynamics and topology of networks while reducing the computational cost of simulations. We show that the semi-metric topology of complex networks yields a natural and algebraically-principled sparsification that outperforms existing methods on those goals. Weighted graphs whose edges represent distances between nodes are \textit{semi-metric} when at least one edge breaks the triangle inequality (transitivity). We first confirm with new experiments that the \textit{metric backbone}—a unique subgraph of all edges that obey the triangle inequality and thus preserve all shortest paths—recovers Susceptible-Infected dynamics over the original non-sparsified graph. This recovery is improved when we remove only those edges that break the triangle inequality significantly, i.e., edges with large semi-metric distortion. Based on these results, we propose the new \textit{semi-metric distortion sparsification} method to progressively sparsify networks in decreasing order of semi-metric distortion. Our method recovers the macro- and micro-level dynamics of epidemic outbreaks better than other methods while also yielding sparser yet connected subgraphs that preserve all shortest paths. Overall, we show that semi-metric distortion overcomes the limitations of edge betweenness in ranking the dynamical relevance of edges not participating in any shortest path, as it quantifies the existence and strength of alternative transmission pathways.}
\end{abstract}

\thispagestyle{empty}


\maketitle

\section{Introduction}

\rev{The advent of network epidemiology in the XXI century~\cite{keeling2005networks,morris2004network} has improved our knowledge about how epidemic outbreaks unfold across real interconnected societies. % The success of network epidemiology in this endeavor builds upon 
The field's increasing relevance for disease control~\cite{metcalf2020mathematical,pagel2022role} has been stimulated by the ability to derive realistic network models that characterize human interactions across multiple scales~\cite{blondel2015survey,stehle2011high}. Indeed, the high resolution characterization of spatiotemporal networks enables actionable interventions to mitigate epidemic outbreaks. For instance, the analysis of mobility networks has shed light on problems as diverse as the risk of importing cases from sources of contagions worldwide~\cite{gilbert2020preparedness,brockmann2013hidden} and the heterogeneous community transmission observed across a given country~\cite{hazarie2021interplay}. Likewise, including networks to refine epidemic models has enabled their use as reliable benchmarks to assess the short-term impact of mitigation and control policies on spreading dynamics~\cite{shea2023multiple}.}

\rev{One of the recurrent problems tackled by network epidemiology is the design of optimal interventions to mitigate an epidemic outbreak while minimally disrupting underlying social, transportation, and trade networks~\cite{meyers2003applying,della2020network}. One such intervention is to focus on nodes, deciding which individuals should be vaccinated first~\cite{rosenblatt2020immunization,mones2018optimizing,wei2022identifying} or where control resources should be prioritized~\cite{reyna2022interconnection,ndeffo2011resource,zhu2021allocating} when facing an epidemic. Another family of interventions targets associations between nodes by reshuffling or removing specific edges 
%or removing specific connections
~\cite{ciaperoni2020relevance} to protect the population from the spread of a circulating pathogen. These strategies are usually driven either by edge-centrality measures such as edge betweeness~\cite{wen2017using,chung2012impact,liang2023effective},
%which might incorporate just structural information of the underlying network, 
%
which typically characterize only the structure of networks, 
or may account 
%for the dynamical state of the system, 
for dynamical properties of edges~\cite{matamalas2018effective}.}

%isolate the agents driving the spread of a virus. For this purpose, node centrality measures help to control network outbreaks by determining, for instance, the individuals who should be vaccinated first when facing an epidemic or the geographical areas that should be prioritized in the spatial distribution of resources. 

\rev{High-resolution network data raises the prospect of digital twins for epidemic forecasting~\cite{de2024challenges}. 
But to be computationally feasible, 
%
%they usually contain many edges which are likely to play a minor role in spreading dynamics across the network, as they represent spurious connections, noise, or sheer redundancy. 
%
\textit{network sparsification} is needed to remove edges that play a minor role in spreading dynamics, without sacrificing important network connectivity and reachability features \cite{de2024challenges,rocha2022feasibility}.
%
%these connections and obtain the smallest subgraph driving epidemic outbreaks and {\em ii})  minimize the computational burden and time necessary to simulate spreading dynamics.
%
Several sparsification methods have been developed over the last couple of decades~\cite{serrano2009extracting,tumminello2005tool}, based on measures of edge importance that focus on different network properties: local node connectivity~\cite{yan2018weight}, statistical impact of edge removal  ~\cite{radicchi2011information,marcaccioli2019polya,gemmetto2017irreducible,
%, impact of edge removal on different network properties 
%(such as the spectrum of their different mathematical representations)~\cite{
imre2020spectrum,bravo2019unifying}, global structure of paths~\cite{spielman2008graph}, or 
redundancy for shortest-path computation 
%subgraph invariance under algebraic closure of probabilistic metric spaces 
in both undirected~\cite{simas2021distance} and directed networks~\cite{costa2023backbone}.
Other methods to reduce the network size use effective renormalization groups~\cite{wilson1974renormalization}, based on either embedding networks on hyperbolic spaces~\cite{garcia2018multiscale} or their Laplacian graph properties~\cite{villegas2023laplacian}.}

The interplay between network sparsification and spreading dynamics has received much less attention than has the design of targeted strategies to control an outbreak. However, mounting evidence in the literature suggests that network sparsification which relies on global information provides better recovery of spreading dynamics than just removing the weakest connections. For instance, sparsifying a network according to the distribution of effective resistance values (which account for the relevance of a given edge within the ensemble of paths that connect its two end nodes) outperforms weights thresholding in preserving both Susceptible-Infected (SI)~\cite{swarup2016identifying} and Susceptible-Infected-Recovered (SIR)~\cite{mercier2022effective} dynamics. Similarly, the study of spreading phenomena through shortest paths in a network has been used to address different problems such as the inference of the source of an outbreak~\cite{tolic2018simulating} and the expected distribution for the arrival times of the pathogen to different locations~\cite{gautreau2007arrival}. Also focusing on shortest paths, a recently published paper by Correia et al.~\cite{Correia:2023contact} shows that the {\em metric backbone}---a unique and algebraically-principled subgraph composed of the (weighted) edges that obey the triangle inequality and thus compose all shortest paths \cite{simas2021distance}
%
%that preserves the entire distribution of shortest-paths of a weighted graph
%
---provides a more solid foundation for network sparsification in epidemic spread than does relying on local information. 

\rev{The metric backbone is a feature of the \textit{semi-metric topology of complex networks} \cite{simas2021distance}. More specifically, weighted graphs can always be isomorphically transformed so that their weights represent distances, allowing analysis via the rich mathematics of (probabilistic) metric spaces \cite{simas2015distance}. This has revealed that the topology of most complex networks derived from real-World data is semi-metric, i.e. the distance between nodes often breaks the triangle inequality \cite{rocha2002semi,simas2015distance}, resulting in much redundancy for computing shortest paths, since only edges that obey the triangle inequality participate in those\cite{simas2021distance}.
However,} determining which features of a given network limit the reconstruction of epidemic outbreaks from the
metric backbone subgraph
%structure of shortest paths in 
remains an open problem, despite previous studies 
%
%on the relevance of the metric backbone
~\cite{Correia:2023contact}.
We tackle this challenge and reveal that improving the reconstruction of spreading dynamics hinges on properties of \textit{semi-metric} edges \cite{simas2015distance, rocha2002semi}. These edges are not included in the metric backbone and, therefore, do not appear in any shortest path in the network.
Specifically, we show that the quality of a reconstruction depends on the interplay between the proportion of semi-metric edges\rev{---which is} complementary to the relative size of the metric backbone---
%, determining the potential transmission pathways missed in the metric backbone, 
%
and their associated \textit{\rev{semi-metric} distortion}, 
%and the semi-metric edge \textit{distortion}
$s^m$~\cite{rocha2002semi,simas2021distance}, a quantification of how much edges break the triangle inequality.
\rev{We show that both these measures characterize how} the semi-metric topology of complex networks \rev{affects spreading dynamics}.
%far they are from the indirect shortest path connecting their nodes. 
%

\rev{Based on these results, we propose and test a new \textit{semi-metric distortion sparsification} (\rev{SMDS}) method, whereby edges are progressively removed in decreasing order of semi-metric distortion. In other words, \rev{SMDS} removes first those edges whose (direct) distance weight is much larger than the length of the shortest (indirect) path between the endpoints they connect, thus breaking the triangular inequality the most \cite{simas2015distance, simas2021distance}. 
%
%Unlike the metric backbone sparsification, which is absolute as it produces a single subgraph of a network, \rev{SMDS} enables to progressively dismantle a network in a deterministic trajectory following the ordering of semi-metric distortion values. 
%
\rev{SMDS} progressively dismantles a network via the (strict, total) ordering of semi-metric distortion values, until the metric backbone subgraph is reached. This algebraically-principled sparsification limit ensures the method preserves all shortest paths of the original network---not only the distribution of shortest path length (reachability) \cite{simas2021distance}. Indeed, sparsification methods that do not preserve the metric backbone necessarily alter the original shortest paths.
We show that \rev{SMDS} outperforms weight and effective resistance thresholding in recovering Susceptible-Infected (SI), Susceptible-Infected-Susceptible (SIS), and Susceptible-Infected-Recovered (SIR) spreading dynamics, for both synthetic and real networks. 
Furthermore,
%it is the only sparsification method that ensures preservation of the full shortest path distribution (reachability), which 
it typically results in sparser subgraphs than the other methods.}

\rev{More generally,  
%studying the structure of shortest paths reveals primary subgraphs for spreading dynamics 
%
the 
%triangular geometry 
semi-metric topology 
of networks, which is induced by distance functions that break the triangle inequality, 
not only defines the edges that are required for shortest paths (the backbone), but also offers a natural and algebraically-principled ranking of the (semi-metric) edges that do not contribute to any shortest path. 
Even though all semi-metric edges have null edge-betweenness, our results demonstrate that their dynamical importance varies widely and is inversely correlated with semi-metric distortion.
While this shows that spreading dynamics does not depend only on shortest paths, it also shows that the edges that are furthest from contributing to any shortest path 
(those that most break the triangle inequality) are dynamically redundant.}

\section{Results}

\subsection{Interplay between \rev{SI} dynamics and the metric backbone}
\label{sec:interplay_si_bbone}
\begin{figure}[t!]
    \centering
    \includegraphics[width=.80\columnwidth]{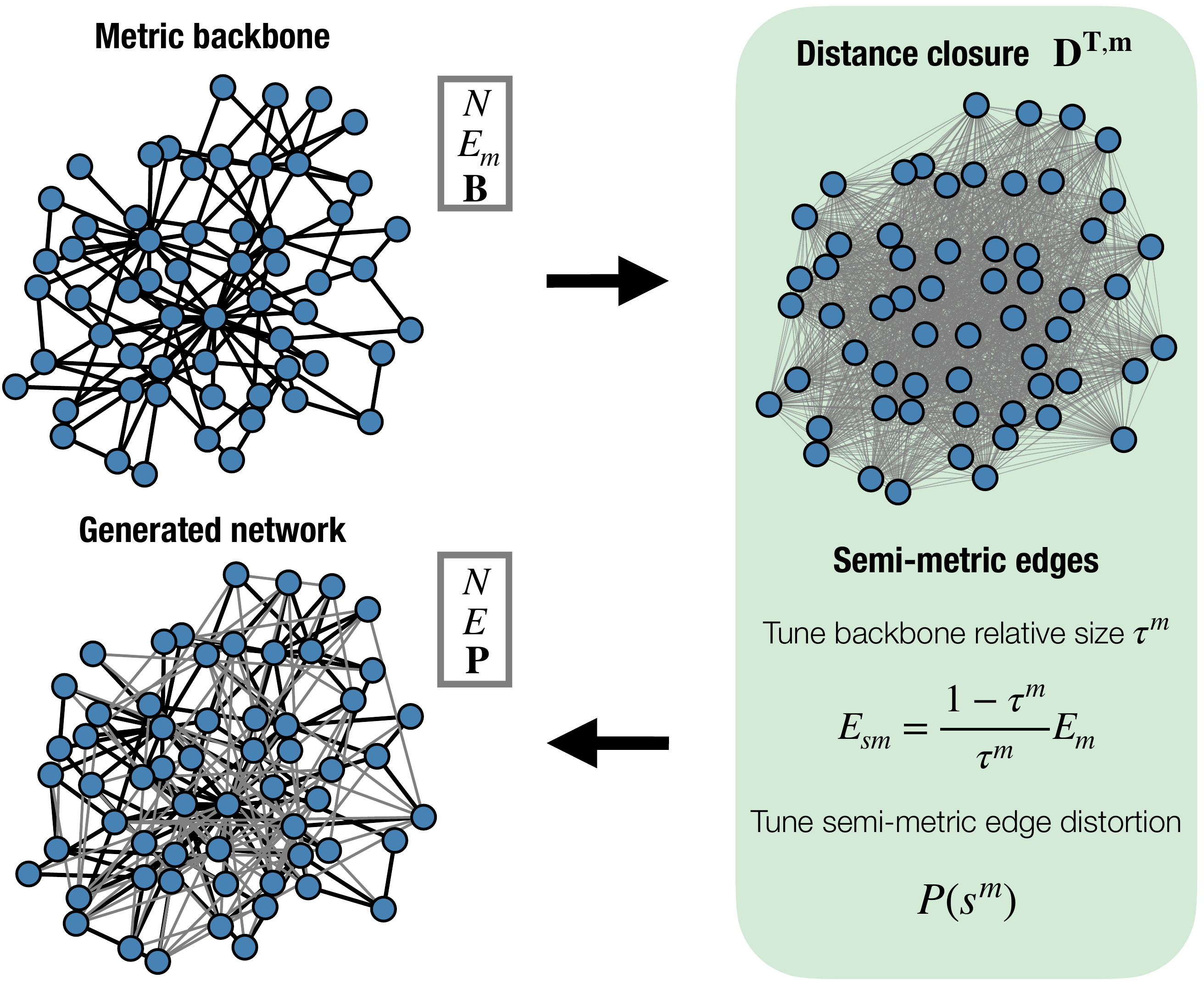}
    \caption{Schematic representation of the construction of synthetic networks with tunable relative size of the backbone $\tau^m$ and semi-metric distortion distribution $P(s^m)$. The first step consists of setting the (connected) metric backbone of a synthetic network. This metric backbone is fully characterized by the number of nodes $N$, the number of edges $E_m$ and the proximity matrix ${\bf B}$ containing the weights of %interactions.
    associations. The next step involves computing the distance closure matrix ${\bf D^{T, m}}$, 
    %a fully-connected graph containing the length of the shortest paths connecting 
     a graph of the same node set $N$, whose edge weights denote the shortest distance (length of shortest path) between every pair of nodes. Finally, we generate a synthetic network by adding $E_{sm}$ edges, whose value is computed according to Eq.~(\ref{eq:semi_metric}) in the Methods section, to obtain a  desired relative backbone size $\tau^m$. The semi-metric distortion values of the added edges $s_{ij}^m$ are drawn from the target distribution $P(s^m)$. Finally, the resulting proximity matrix ${\bf P}$ is obtained by applying Eqs.~(\ref{eq:s}) and (\ref{eq:d}) in Methods section to each $s^m_{ij}$ value.}
    \label{fig1:synthetic_creation}
\end{figure}

We first explore different network features that potentially shape the reconstruction of epidemic outbreaks from the metric backbone. 
To address this challenge, we propose a new method to construct synthetic networks where we can tune the relative size of the metric backbone $\tau^m$ and the distribution of semi-metric distortion values $P(s^m)$, as shown in Figure~\ref{fig1:synthetic_creation}. We refer the reader to the Methods subsection \rev{``}Metric backbone and semi-metric distortion\rev{''} for a complete explanation of the theoretical foundations of the metric backbone and the associated semi-metric distortion parameter of an edge. Starting from a metric backbone of $N$ nodes and $E_m$ edges, our method adds semi-metric edges to build a network of $N$ nodes and $E$ edges, whose strength of associations are quantified by the proximity matrix ${\bf P}$. A detailed description of our method to construct synthetic networks is provided in the Methods subsection \rev{``}Construction of synthetic networks\rev{''}. 

To focus on the role of the aforementioned features, $\tau^m$ and $P(s^m)$, while preserving other empirically relevant structures, the synthetic networks are built from the backbone of a real network. Specifically, we utilize the backbone of a contact network between elementary school students (kindergarten to sixth grade) in Utah (USA) \cite{toth2015role, Correia:2023contact}.  This metric backbone is composed of $N=339$ nodes and $E_m=1128$ metric edges, with a heterogeneous distribution of the proximity values (available in Figure S1 in the Supplementary text). 
Moreover, for the semi-metric edges, their semi-metric distortion values are drawn from a log-normal distribution via $s^m=1+r$ where  $r\sim Lognormal(\mu,\sigma)$. Note that this probability function is widespread across different empirical networks~\cite{simas2021distance,Correia:2023contact}. 
Throughout the manuscript, we fix $\sigma=1$ and modify the $\mu$ value to study the effect of semi-metric edges' relevance with respect to those included in the metric backbone. 

To assess the extent to which the metric backbone can reconstruct epidemic outbreaks, we introduce a single infectious seed, i.e. a single individual initially infected, and run a SI dynamics, which constitutes the simplest framework to simulate spreading phenomena in networks. As detailed in the Methods subsection \rev{``}SI model\rev{''}, epidemic outbreaks in the SI model are fully described by the distribution of times of infection for the different nodes across the network. Based on this fact, we assume that the characteristic time scale of SI dynamics in a given network is the time at which half of the population is reached by the outbreak, denoted in here by $t_{1/2}$. Therefore, our research question narrows down to determine whether the characteristic time scale of spreading phenomena in the metric backbone \rev{${\bf B}$}, denoted by \rev{$t^{B}_{1/2}$}, captures the corresponding measure in the entire network. \rev{Note that, in the following comparative measures, the superscript refers to the sparsification method while the subscript to the time of comparison. We define the ratio comparing the times at which half of the population is reached in the backbone and the original network, }\rev{$\xi^{B}_{1/2}$}, as:
%
%Further details on the SI model and the chosen parameters can be found in the Methods section. To quantify the extent to which the metric backbone preserves SI dynamics given an initial infectious seed, we compute the ratio between the former quantities, yielding:
\rev{\begin{equation}
\xi^{B}_{1/2}=\frac{t_{1/2}^{B}}{t_{1/2}}\ .
\end{equation}}
In absence of stochastic fluctuations, the aforementioned ratio fulfills \rev{$\xi^{B}_{1/2} \geq 1$}, as the metric backbone always removes potential transmission pathways for the virus existing in the original network. In terms of performance, the closer this ratio gets to \rev{$\xi^{B}_{1/2}=1$}, the more faithful the information provided by the metric backbone is about the dynamics in the entire network. %We obtain the former ratios for $50$ different initial seeds to smooth out possible biases introduced by the origin of the outbreak in our analysis. (\rev{Maybe we can move some of this information to the Methods section, although I think it is crucial to understand the figures and therefore I will leave it here})

\begin{figure}[t!]
    \centering
    \includegraphics[width=.80\columnwidth]{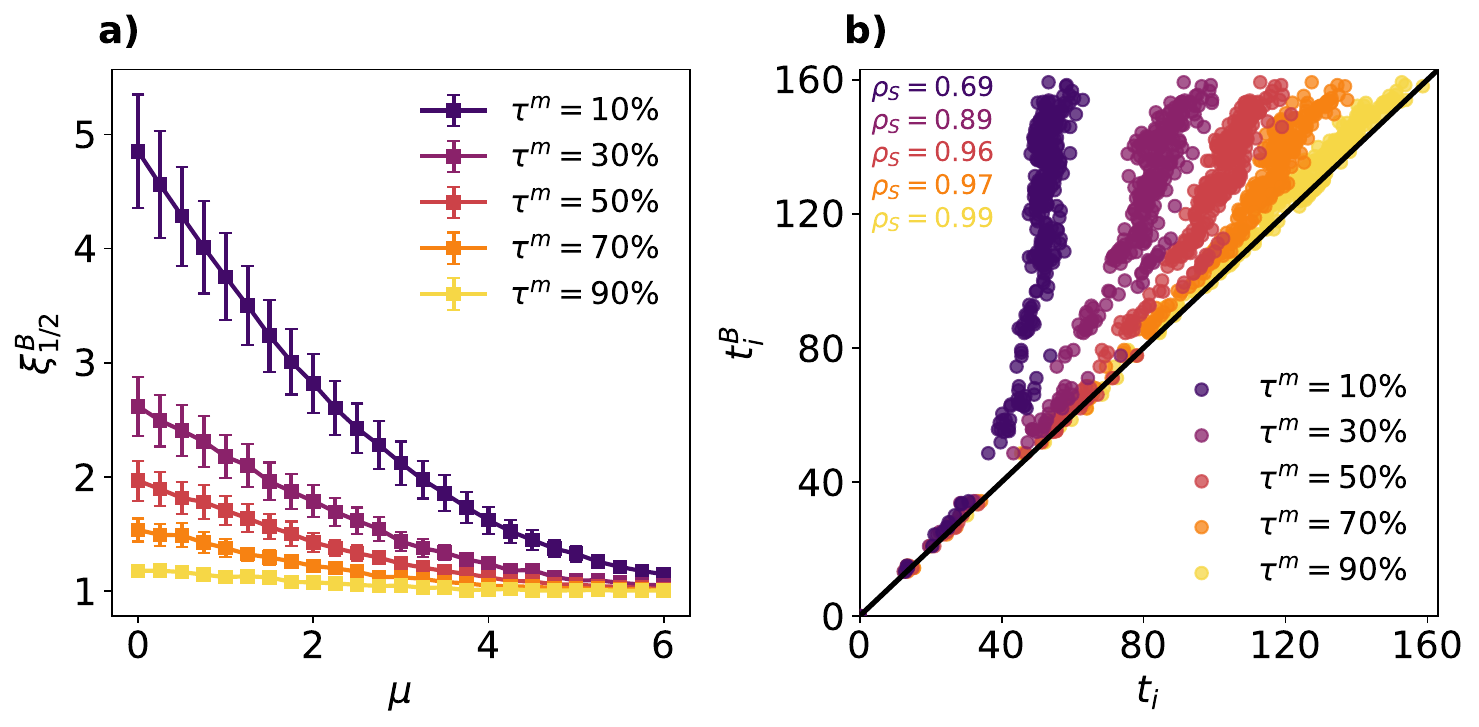}
    \caption{Panel a): Ratio between the time for the disease to reach half the population in the metric backbone and in the constructed network \rev{$\xi_{1/2}^{B}$}  as a function of the parameter $\mu$, i.e. the logarithm of the median of the semi-metric distortion distribution, and the relative size of the backbone $\tau^m$ (color code). For a given seed, we compute the time to reach half the population $t_{1/2}$ as the median value observed across $200$ realizations of the SI dynamics. Dots show the average and error bars represent the standard deviation of the ratios obtained for $50$ different infectious seeds.   Panel b): Average time of infection of each individual $i$ in the metric backbone \rev{$t_i^{B}$} as a function of its average time of infection in the entire network $t_i$ for different values of the relative size of the backbone $\tau^m$ (color code). These results are obtained by averaging $200$ realizations starting from the same initial seed. In both panels, we set $\beta=0.5$ for the SI dynamics, fix $\sigma=1$ in the semi-metric distortion distribution and consider the metric backbone ${\bf B}$ of a network capturing face-to-face interactions in a elementary school in the US~\cite{toth2015role}.} 
    \label{fig2}
\end{figure}

Figure ~\ref{fig2}a represents the ratio \rev{$\xi^{B}_{1/2}$} as a function of the semi-metric edges relevance, governed by $\mu$, and the relative size of the backbone $\tau^m$. For large $\mu$ values, semi-metric edges are highly dynamically redundant in comparison with the metric ones as \rev{$\xi^{B}_{1/2} \simeq 1$}regardless of the backbone relative size. Therefore, their removal hardly has any influence on the spreading. As edges distances become closer to the shortest path lengths, we observe a critical value $\mu_c$ for each $\tau^m$ value below which the spreading dynamics gets slower in the metric backbone ($\xi^{B}_{1/2} > 1$). Interestingly, this critical value $\mu_c$ increases as the size of the backbone decreases. The latter results imply that the performance of the metric backbone is determined by the interplay between both the potential transmission pathways pruned during the sparsification process (governed by $\tau^m$) and their relevance with respect to those kept in the metric backbone (the semi-metric distortion, governed by $\mu$). 

The previous results show that considering the metric backbone as the underlying contact structure might induce global delays in the spreading dynamics. Nonetheless, even in those scenarios, the information obtained from this subgraph can be relevant for disease control if the metric backbone allows us to faithfully rank the different nodes according to their expected time of infection. To check that, we randomly place a single infectious seed in the network and study how the distribution of the individual infection times varies as we alter the properties of the metric backbone. In particular, we fix $\mu=2$ and explore the role of the relative size of the backbone $\tau^m$ in the microscopic reconstruction of outbreaks. Figure ~\ref{fig2}b shows that, even when the metric backbone represents just $10\%$ of the edges of the network, it qualitatively captures the epidemic trajectory across the population, as shown by the high Spearman correlation between the distributions obtained for both the network and its backbone ($\rho_S=0.69$, $p<0.001$). As expected, the microscopic recovery of the epidemic trajectory is also enhanced as $\tau^m$ increases and more transmission pathways are captured in the metric backbone.

\subsection{\rev{Semi-Metric Distortion Sparsification} in synthetic networks}
\label{sec:sparsification}

\begin{figure}[t!]
    \centering
    \includegraphics[width=.80\columnwidth]{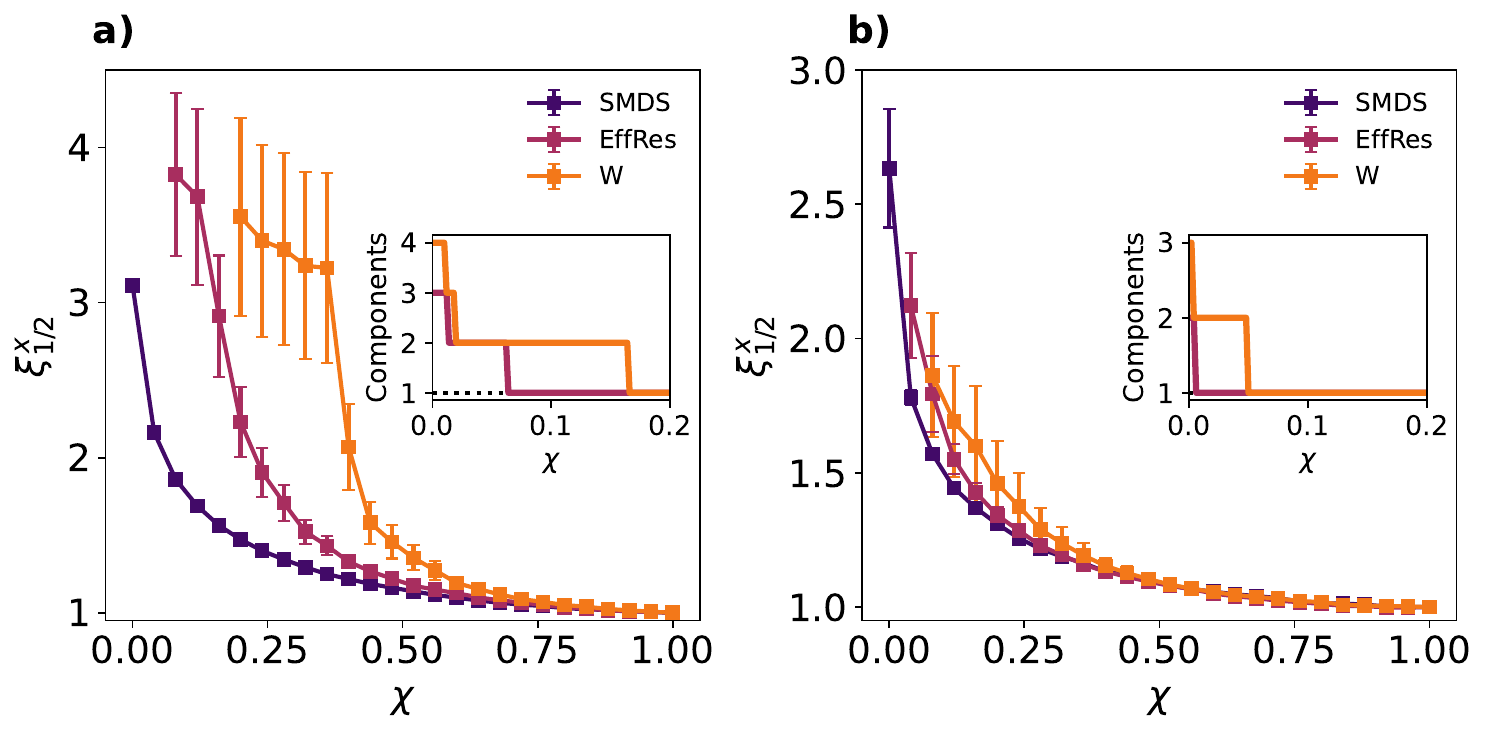}
    \caption{Ratio between the time for the disease to reach half the population in the sparsified network and in the original synthetic network $\xi^{x}_{1/2}$ as function of the parameter governing the size of the sparsified network $\chi$ for each sparsification method $x$ (color code). $\chi$ ranges from $\chi=1$, corresponding to not removing any edge, to $\chi=0$ where $E_{sm}$ edges are removed. Three different sparsification methods are compared: targeting edges with highest semi-metric distortion values (\rev{SMDS}), lowest effective resistance values (EffRes) or lowest proximity values (W). The details of the simulations to obtain the ratios are the ones described for Figure~\ref{fig2}A. EffRes and W curves are interrupted when breaking the largest connected component of the network. Insets: Number of components in the network as a function of the edges removed $\chi$ and the sparsification method (color code). The horizontal dashed line represents the \rev{SMDS}, which always preserves the largest connected component. The original synthetic networks are constructed considering $\tau^m=0.1$ and $\mu=1$ (Panel a) or $\mu=2$ (Panel b) respectively.} 
    \label{fig3}
\end{figure}

Our findings indicate that semi-metric edges with large $s^m$ values are dynamically redundant, as their removal has negligible effects on SI dynamics. Motivated by this result, here we propose \rev{the \textit{semi-metric distortion sparsification} (SMDS) method, where} we sort the edges $(i,j)$ according to their associated distortion $s_{ij}^m$ and \rev{progressively} remove those with highest values until matching the desired size of the sparsified configuration. 
%Note that this method is deterministic, hereinafter referred to as \rev{SMDS}, yielding a single network trajectory from the entire network to its associated metric backbone.
We compare the method proposed here with two other sparsification schemes relying on different edges properties: \textit{weights thresholding} and \textit{effective resistance thresholding}. On the one hand, weights thresholding relies on local information, aiming at removing the weakest connections through which transmission of the virus is very unlikely. On the other hand, effective resistance thresholding penalizes path redundancy, as small effective resistance values identify those direct edges connecting nodes which can also exchange information through many other indirect paths. More details on the computation of the effective resistance associated to each edge can be found in the Methods subsection \rev{``}Effective resistance\rev{''}. 

To assess and compare performance, we compute the ratio $\xi^{x}_{1/2} (\chi)=t^{x}_{1/2} (\chi)/t_{1/2}$ for each subgraph obtained after removing edges according to each sparsification method $x$.
For the \rev{SMDS}, 
%thresholding
%
$\chi$ denotes the fraction of semi-metric edges included in the graph, i.e. $E(\chi) = E_m + \chi E_{sm}$. Therefore, $\chi=1$ corresponds to the entire graph whereas $\chi=0$ corresponds to the metric backbone subgraph. For the other two methods, the sparsified networks comprise $E(\chi)$ edges which are chosen by thresholding proximity weights $p_{ij}$ or effective resistance values $p^R_{ij}$, defined according to Eq.(~\ref{eq:p}) and Eq.(~\ref{eq:r}) in the Methods section, respectively.

Figure~\ref{fig3}a shows the comparison of the three sparsification methods in a synthetic network with $\tau^m=0.1$ and $\mu=1$. We observe that \rev{SMDS} outperforms the other two methods in both preserving SI dynamics and keeping the connectedness of the network.
Specifically, $\xi_{1/2}$  for \rev{SMDS} remains closer to $1$,
%
%implying that accounting for the edge relevance with respect to their associated shortest path during sparsification helps reconstruct the spreading dynamics. 
%
showing that edge relevance for SI dynamics is more correlated with their semi-metric distortion than their proximity weights or effective resistance. 
This demonstrates that even semi-metric edges that do not contribute to any shortest path (they have zero betweenness), are more relevant to SI dynamics the less they break the triangle inequality. 
In other words, those edges that are nearer to being necessary for computing shortest paths (low semi-metric distortion) play a much more important role in SI dynamics than the edges that are far from contributing to shortest paths (large semi-metric distortion). 

It is also important to notice that \rev{SMDS} does not target any metric edge by definition and therefore guarantees the preservation of the shortest path connecting every pair of nodes, preserving both connectivity and shortest distances. In contrast, the other two methods eventually dismantle the largest connected component, breaking the network into different subgraphs. Note that this phenomenon is more pronounced for thresholding by weights than by effective resistance as the latter harnesses global information and makes the isolation of individual nodes rarer. 

Interestingly, increasing $\mu$ in the synthetic network reduces the differences between the three methods, as shown in Figure~\ref{fig3}b for networks constructed with $\mu=2$. In this case, the synthetic networks have many more semi-metric edges with a large distortion, 
%are far from the shortest path, 
which increases the chances that both other methods do not target the metric backbone when thresholding by proximity weights or effective resistance values, as demonstrated in Figure S2 in the Supplementary text. 
This supports the role of the metric backbone as a primary subgraph sustaining the spread of diseases~\cite{Correia:2023contact}, but it also newly reveals that edges with larger semi-metric distortion rarely contribute to spreading dynamics (notice difference in scale for $\xi^{x}_{1/2}$ between panels a and b in Figure~\ref{fig3}) .

\subsection{\rev{Semi-Metric Distortion Sparsification} in empirical networks}
\label{sec:real_nets}

\begin{figure}[t!]
\centering
\includegraphics[width=.80\columnwidth]{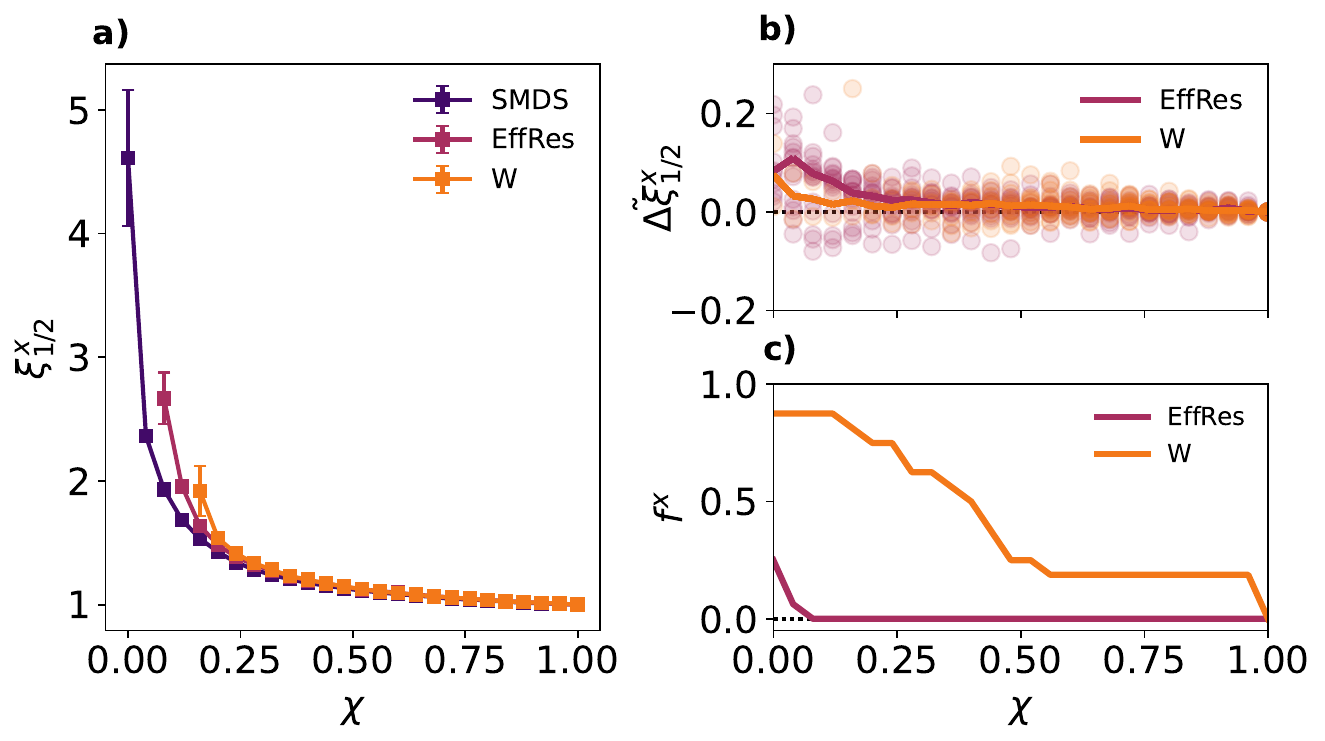}
\caption{Panel a): Ratio between the time for the disease to reach half the population in the sparsified network and in the original synthetic network $\xi^{x}_{1/2}$ as a function of the parameter $\chi$, governing the edges removed from the network, and the sparsification method $x$ (color code). The contact network corresponds to face-to-face interactions recorded in an elementary school students (kindergarten to sixth grade) in Utah (USA)~\cite{toth2015role}. The details of the simulations to obtain the ratios are the ones described for Figure~\ref{fig2}A. Panel b): Distribution of the relative difference $\tilde{\Delta\xi}^x_{1/2}$ between the mean ratio associated to each sparsification method $x$ (color code) and 
%the one following 
\rev{SMDS} as a function of the size of the sparsified network $\chi$, for all emprical networks. Positive $\tilde{\Delta\xi}^x_{1/2}$ values reveal a better performance of \rev{SMDS} with respect to the sparsification method $x$. Dots represent each of the 16 networks included in the dataset under study whereas solid lines show the median $\tilde{\Delta\xi}_{1/2}^x$ value observed across networks for each sparsification method $x$. Panel c): Fraction of disconnected networks in the dataset of real networks $f^x$ as a function of the size of the sparsified network $\chi$ and the sparsification method $x$ (color code). In panels b)-c), the horizontal dashed line represents the values obtained for the \rev{SMDS}.}
\label{fig4}
\end{figure}

We also study 16 networks built from biological, social and transportation data, as detailed in the Methods subsection \rev{``}Empirical Networks\rev{''}.
%We refer the reader to the Methods section for further information about the construction of these networks, whose structural features are summarized in Table S1 in the Supplementary text.
%
Figure~\ref{fig4}a depicts a comparison between the three sparsification methods in regards to recovering SI dynamics for the case the US elementary school social contact network---specifically, time to infect half of the population. It is clear that \rev{SMDS} presents the same advantages observed in the case of synthetic networks, i.e. better recovery of SI dynamics overall and no disruption of the largest connected component of the network for greater sparsification. 
A similar analysis for all other networks is provided in Figure S3 of the Supplementary text, leading to the same overall results.

To quantify the superior recovery of SI dynamics observed for \rev{SMDS} across empirical networks, we compute the relative differences between the mean ratios observed for the \rev{SMDS} and each other method $x$, as a function of the size of the sparsified subgraph $\chi$: 
\begin{equation}
    \tilde{\Delta\xi}^x_{1/2} (\chi) =\frac{\xi^x_{1/2}(\chi)-\xi^{\rev{SMDS}}_{1/2}(\chi)}{\xi^{\rev{SMDS}}_{1/2}(\chi)}\ ,
\end{equation}
where, similarly to the synthetic network study (\S \ref{sec:sparsification}), $x$ stands for thresholding by either proximity weights or effective resistance values.
Therefore, when $\tilde{\Delta\xi}^x_{1/2}>0$, \rev{SMDS} outperforms method $x$.

Similarly to what was observed in the synthetic network study, Figure~\ref{fig4}a shows little-to-no differences across sparsification methods when removing few edges from the network, as $\xi^{x}_{1/2} \simeq 1$ when $\chi\simeq 1$ regardless of the method considered. Nonetheless, as more edges get pruned, \rev{SMDS} typically outperforms the other two methods as shown by the positive median of the distribution of $\tilde{\Delta\xi}^x_{1/2}$ values obtained for the different networks in the dataset (Figure~\ref{fig4}b). However, there are some networks for which $\tilde{\Delta\xi}^x_{1/2}<0$, which means that the other methods can outperform \rev{SMDS} in preserving SI dynamics for some empirical networks, such as the Human Connectome network \rev{(see Fig. S3 for the comparison of the methods for each network)}.
To further support our findings, Figure S4 in the Supplementary text presents a comparison between the three methods for different stages of an epidemic outbreak. There we observe that \rev{SMDS} generally allows for a better construction of the dynamics except for early stages of the epidemic spreading, when only $10\%$ of the population has been infected. In that case, the set of infected individuals reflects the composition of very localized infection around the seed node, thus making the global shortest-path information included by semi-metric distortion values less relevant. 

\rev{Apart from recovering spreading dynamics, a sparsified network should preserve the 
%largest connected component 
connectivity
of the original network to ensure its functioning. For each sparsification method $x$,} we represent in Figure~\ref{fig4}c the fraction $f^x$  of empirical networks that become disconnected as a function of the size of their sparsified subgraphs. We observe that the empirical networks are quite vulnerable to thresholding by proximity weights, as most of the networks in our dataset eventually break into different components before reaching the size of the metric backbone subgraph.
While thresholding by effective resistance is more resilient (likely due to the global information encoded in the effective resistance values), 4 out of 16 networks end up breaking into separate components when removing the same number of (but not the same) edges that get removed to obtain the metric backbone ($\chi=0$).
Figure S5 in the Supplementary text further shows the number of components as a function of the size $\chi$ for each of the different networks studied. 
%, showing that the way in which these sparsification methods dismantle networks is not universal but highly dependent on the specific architecture of each graph.

\rev{To complete our analysis, we further consider how different sparsification methods preserve epidemic outbreaks under Susceptible-Infected-Susceptible (SIS) and Susceptible-Infected-Recovered (SIR) dynamics, which are more complex compartmental models. The detailed results are shown in the Supplementary Materials. In both cases, we set the epidemiological parameters to ensure the existence of a widespread epidemic state in all empirical networks. 
%
%Figure S6 shows that weights thresholding, when not breaking the network, preserves slightly better the global epidemic state in the SIS dynamics while similar performances of \rev{SMDS} and effective resistance thresholding are observed across the entire range of sparsified networks. 
%
Figure S6 shows that \rev{SMDS} and effective resistance thresholding lead to similar macroscopic SIS dynamics performance, reaching a similar global epidemic state for the same amount of sparsification. However, in a majority of networks and across sparsification levels, \rev{SMDS} reaches a fraction of infected individuals at steady state closer to that of the original network, without ever disconnecting it (which effective resistance can do, as shown in Figures S4 and S5).
Weight thresholding frequently breaks the original networks into components even for modest amounts of sparsification. Though, when this sparsification does not break a network, it preserves slightly better the global epidemic state of the original network. 
%
%Nonetheless, \rev{SMDS} outperfoms effective resistance thresholding in preserving the ranking of nodes according to their individual probabilities of infection, thus providing a better microscopic picture of the epidemic outbreak (Figure S7).
%
When it comes to microscopic SIS dynamics of an epidemic outbreak, \rev{SMDS} outperfoms effective resistance thresholding in preserving the ranking of nodes according to their individual probabilities of infection (Figure S7).
Similar results are observed for macroscopic and microsopic SIR dynamics in Figures S8 and S9, respectively. These results, along with those reported for the SI dynamics, demonstrate that \rev{SMDS} is 
%a suitable 
an excellent sparsification method to generate subgraphs preserving spreading dynamics and the connectedness of complex networks.}  

\section{Discussion}
\label{sec:conclusions}

Sophisticated data gathering techniques now enable the generation of large-scale networks that represent the architecture of relationships spanning multiple scales in nature, ranging from biochemical interactions to social contacts and transportation flows, all of which greatly contribute to the spread of disease \cite{pastor2015epidemic, balcan2009multiscale}.
Despite much scientific progress in understanding how epidemics spread via such networks, the resulting models usually include redundant associations that do not contribute to their dynamics ~\cite{marques2013canalization, gates2021effective, mercier2022effective, Correia:2023contact}. 
These redundant associations not only enlarge the networks and their computational analysis \cite{simas2015distance,simas2021distance}, but also may obfuscate their topology and causal interaction structure leading to erroneous predictions \cite{gates2016control, costa2023effective, manicka2022effective}.
Network sparsification methods are thus essential to leverage the high spatiotemporal resolution of current datasets toward effective and actionable models such as digital twins \cite{de2024challenges}.

Here we investigate whether the \rev{semi-metric topology of weighted graphs} provides reliable guidance for sparsifying a network while preserving spreading dynamics. To do so, first we study the 
%topological features of weighted graphs that determine the 
reconstruction of SI outbreaks from the metric backbone, which is composed of the subset of edges that obey the triangle inequality 
%underlying a standard metric space 
\cite{simas2021distance}.
While typically much smaller than the original network, \rev{the edges of the metric backbone are} necessary and sufficient to compute all shortest paths in the original network. 
Our results show that it is indeed a primary subgraph for transmission dynamics. This is true for both the macro-dynamics of the overall times to infection, as previously suggested \cite{Correia:2023contact}, and also for the micro-level dynamics captured by the time at which individuals get infected (see Figure \ref{fig2}b and \S \ref{sec:sparsification}).
Specifically, even though the metric backbone induces a global delay in the spreading dynamics, it faithfully classifies the order in which individuals get exposed to the outbreak, and thus provides useful guidance for prioritizing control and mitigation strategies to disrupt epidemic spreading.

Interestingly, our results reveal that the impact \rev{on dynamics of the (semi-metric)} edges that are not included in the metric backbone varies considerably.
Indeed, effective reconstruction of outbreaks is shaped by a nontrivial interplay between semi-metric edge quantity and quality, by which we mean how much they break the triangle inequality, given by the semi-metric distortion parameter $s^m$ (eq. (\ref{eq:s}), Methods section).
Edges with $s^m \gg 1$ are very far from contributing to any shortest path, whereas those with $s^m \approx 1$ link nodes with a distance almost the same as the shortest path (\S \ref{sec:sparsification})---in other words, the ``cost'' of connecting two nodes directly via a low distortion edge is little more than via the (slightly shorter) indirect shortest path.

Our results demonstrate that removing high distortion edges hardly affects the dynamics regardless of the relative size of the backbone. In contrast, semi-metric edges with small distortion are important for recovering spreading dynamics. 
%
%\rev{This means that even when smaller sparsification is possible, such as the ultrametric backbone \cite{simas2021distance, costa2023backbone} for example, it comes at the expense of information retrieval.}
%
When there are more of these semi-metric edges, the size of the metric backbone becomes much more relevant for recovering the spreading dynamics accurately, whereby smaller backbones on their own do not recover dynamics as effectively as larger backbones (see Figures \ref{fig2}a, Figure \ref{fig3}, and \S \ref{sec:sparsification}).
This suggests that availability of low distortion edges makes epidemic outbreaks robust by providing ``near shortest'' transmission alternatives to shortest paths, whereas high distortion edges very rarely provide alternative paths in spreading dynamics and can be safely removed.
\rev{Notice that it is possible to remove edges from the metric backbone itself and still preserve the distribution of shortest distances (or path lengths) of the original network \cite{simas2021distance}. This happens when there are alternative shortest paths with the exact same length between some node pairs. Importantly, the edges of all alternative shortest paths are kept in the metric backbone since they all obey the triangle inequality (because it is not a strict inequality). Our results show that those alternative equivalent paths should indeed be kept in sparsified networks as the metric backbone does, because they are not redundant for spreading dynamics as they have small semi-metric distortion.}

These results indicate that measuring the \rev{semi-metric topology of complex networks, induced by distance functions that break} the triangle inequality, is crucial for accurate reconstruction of spreading phenomena from sparsified networks. 
%Indeed, the metric backbone and the associated semi-metric distortion parameter provide the most effective edge relevance rank we studied in comparison to other sparsification methods in both synthetic and empirical networks.
%
We thus developed and evaluated the \rev{SMDS} to progressively sparsify weighted graphs by first removing the edges with the largest semi-metric distortion. 
\rev{Complementary to the unique sparsification given by identifying the metric backbone, \rev{SMDS} can progressively reduce the entire network until its metric backbone, allowing for more or less aggressive sparsification.}
We compare \rev{SMDS} with two sparsification methods by thresholding (neither of which has a principled limit): 1) using proximity weights, a local sparsification method that targets the weakest connections, and 2) using effective resistance, a global sparsification method that penalizes the existence of multiple paths connecting two nodes.
We show that \rev{SMDS} not only yields a more accurate reconstruction of macro and micro spreading dynamics, but also ensures network functionality by not breaking the network into disconnected components even while providing greater sparsification and preserving all shortest paths. 
More broadly,
%the prominent role of semi-metric edges, 
%the metric topology of weighted graphs 
\rev{our study} reveals that the distribution of semi-metric distortion values---especially the existence of many edges with small distortion---does play an important role in spreading dynamics.
Because all semi-metric edges have zero edge betweenness, this result implies that new edge centrality measures that incorporate information about the \rev{semi-metric topology of weighted graphs} would be useful to better capture edge importance.  

Taken together, our findings \rev{demonstrate that the semi-metric topology of complex networks} provides an algebraically-principled approach to network sparsification 
%\rev{(both progressive as well as absolute) }
that is more effective than alternatives in recovering spreading dynamics.
\rev{It should be noted that our results are restricted to widespread epidemic states. Under SI dynamics, the emergence of a global infected configuration is guaranteed and shortest paths appear as a natural driver for contagion processes. Likewise, our analysis of SIS and SIR dynamics is focused on outbreaks affecting a large fraction of the population. However, these models} usually display localized epidemic outbreaks~\cite{silva2021dissecting,cota2016griffiths}, in which the 
%prevalence 
\rev{importance of global 
%information 
communication
over local 
%weights 
connectivity }for network sparsification remains an open question. Additionally, shortest paths have been reported to not capture some \rev{network diffusion processes, such as random walker dynamics, which can flow via} longer but more likely network bypasses 
%governing the flow of information across the network
~\cite{estrada2023network,lella2020communicability}.
Far from being limitations, these issues call for extending our study beyond shortest paths \rev{defined by distance functions constrained by the standard triangle inequality. 
We should consider general distance backbones induced by a generalized triangle inequality that can consider any measure of path length (not just the sum of distance weights) to characterize indirect transmission cost measures for other forms of spreading dynamics \cite{simas2021distance}.}
%(\S \ref{sec:backbone_method})~\cite{simas2015distance,simas2021distance}.
%
We are confident that both the methodology presented here and our results will motivate further study of the role of backbone subgraphs involved in driving and controlling network dynamics. 

\section{Methods}

\subsection{Metric backbone and semi-metric distortion} 
\label{sec:backbone_method}

%Here we give a brief overview about the theoretical foundations of the metric backbone and the semi-metric edge distortion.
Let us assume that we have a system of $N$ individuals whose contact associations are quantified by a proximity matrix ${\bf P}$, with entries $p_{ij}$ denoting the strength of association between nodes (individuals) $i$ and $j$---typically a normalized measure of how often the two individuals were together \cite{Correia:2023contact}. Obtaining the metric backbone involves the computation of shortest paths in the network; thus, we need a map between the proximity matrix ${\bf P}$ and \rev{a} distance matrix ${\bf D}$, whose entries, $d_{ij}$, represent a notion of distance between nodes $i$ and $j$, such that a small value denotes a strong association. 
\rev{Furthermore, this map should not affect the topology of ${\bf P}$, requiring it to be an isomorphism, such as}:
\begin{equation}
    d_{ij} = \frac{1}{p_{ij}}-1 \ ,
\label{eq:d}
\end{equation}
with $d_{ij}\to\infty$ in case two individuals are never together~\cite{simas2015distance}. 
\rev{This isomorphic relation is an important feature of the methodology, as it allows us to convert all key features of the semi-metric analysis to any weighted graph \cite{simas2021distance}. Thus, concepts such as shortest paths, the metric backbone, and semi-metric distortion introduced below are guaranteed to exist in weighted graphs whose edges denote proximity or strength, instead of distance or dissimilarity \cite{Correia:2023contact}. This easy conversion ensures that original meaning of edge weights is preserved, adding to the explainability power of the method.}

Once the direct distances between nodes are defined, we can 
%compute the distance of a given 
consider paths that connect nodes $i$ and $j$ indirectly: $\Gamma=\qty{ i,k_1, \dots, k_n,j }$, where $n$ is the number of intermediate nodes in the path. In the case of the metric backbone, the length of a path is computed as 

\begin{equation}
d_{ij}^m(\Gamma)=d_{ik_1}+d_{k_1k_2} + \dots + d_{k_n j} \ .
\label{eq:path_distance}
\end{equation}

\noindent \rev{Note that we use the superscript $m$ to indicate that all the measures in this manuscript are related to the metric algebraic space.} Other choices for computing path length are possible, leading to generalized distance backbones~\cite{simas2021distance}, which are to be considered in future work.
The shortest path \rev{metric} length connecting both nodes (or shortest indirect distance) is in turn computed as
%$d^{T, m}_{ij}=\min_{\Gamma} \qty( \qty{ d_{ij}(\Gamma) } )$, 
$d^{T, m}_{ij}=\min \qty( \qty{ d_{ij}(\Gamma) } )$.
%
%,which by construction satisfies $d_{ij}^{T, m} \leq d_{ij}$.
%
Computing the shortest path \rev{metric} length for all node pairs in ${\bf D}$ is known as the \textit{all-Pairs Shortest Path problem} \cite{zwick2002all} and results in
%
%The collection of shortest path distances is encoded in 
%
the metric closure matrix, denoted by ${\bf D^{T, m}}$ \cite{simas2015distance}.
Comparing the entries of this matrix with the corresponding finite entries of ${\bf D}$ reveals which edges obey or break \rev{the} triangle inequality for any indirect path (via any number of indirect nodes): $d_{ij}  \leq d_{ij}^{T, m}, \forall_{ d_{ij} < \infty}$. In the first case, there are no shorter indirect paths, and the edge is referred to as {\it metric} and denoted and defined as $b_{ij} \equiv d_{ij}^{T,m} = d_{ij}$. However, in real-\rev{w}orld networks there are typically many {\it semi-metric} edges that break this triangle inequality, observing $d_{ij} > d_{ij}^{T, m}$ instead \cite{rocha2002semi}.
Importantly, only the metric edges are necessary to compute all shortest paths (and sufficient to compute all shortest indirect distances), and thus define 
the \textit{metric backbone} ${\bf B}$, whose \textit{relative size} is given by 
\begin{equation}
    %\tau^m = \frac{\abs{\{d_{ij} < \infty : d_{ij} = d_{ij}^T\}}}{\abs{\{d_{ij} < \infty\}}},
        \tau^m = \frac{\abs{\{b_{ij}\}}}{\abs{\{d_{ij} < \infty\}}},
    \label{eq:tau}
\end{equation}
measuring the proportion of edges kept in this subgraph \cite{simas2021distance}. 
For unweighted graphs, $\tau^m=1$, as all the direct edges represent the shortest path between their nodes. In contrast, for weighted graphs $\tau^m \leq 1$,  depending on both the weights distribution and the specific structure of paths in each graph.
%
%The metric backbone is just one of the distance backbones which can be extracted from the distance matrix ${\bf D}$, according to how the length of a path is measured and how the different paths are combined. While in this work we are exclusively interested in the metric backbone,  we refer the reader to~\cite{simas2021distance} for a more exhaustive exploration of other backbones and their properties.
%
Unless otherwise noticed, in this article when we refer to the triangle inequality, we mean the generalized case above which considers all possible indirect paths of any length (not simply directly connected triangles).

Since semi-metric edges do not participate in any shortest paths, they all have null edge betwe\rev{e}nness (and associated centrality measure). 
However, they vary widely in their \textit{semi-metric distortion} parameter given by:
%
%\footnote{The subscripts $T, m$ indicates that the positive real distances are transitive with respect to addition.}.
%% From here to line 87, change order?

\begin{equation}
    s_{ij}^m = \frac{d_{ij}}{d^{T, m}_{ij}}\ ,
    \label{eq:s}
\end{equation}
%From all the edges participating in shortest paths, we can construct the metric backbone ${\bf B}$, whose elements $b_{ij}$ are given by:
%
which quantifies how much the edge between nodes $i$ and $j$ breaks the triangle inequality---in other words, how much shorter is the indirect distance (shortest path \rev{metric} length) than the direct distance between them.
For metric edges in the backbone, since the triangle inequality is satisfied, we naturally have $s_{ij}^m=1$, while for semi-metric edges we have $s_{ij}^m>1$.

\rev{W}e refer the reader to~\cite{simas2021distance} for a more exhaustive definition and study of the metric and other distance backbones, including the observed values of these parameters in networks across many domains.

\subsection{Construction of synthetic networks}
\label{sec:synthetic_construction}
The first step to construct the synthetic networks used in this manuscript is to fix their metric backbone. To do so, we consider an undirected weighted network with $N$ nodes and $E_m$ edges, whose proximity values are captured in the matrix ${\bf B}$. This initial subgraph has to satisfy all the constraints characterizing the metric backbone of a given network. Namely, it must have a single connected component and all its edges must be metric, i.e. they must represent the shortest path connecting their nodes. The next step involves mapping the proximity values to distances via Eq.~(\ref{eq:d}) and computing the metric closure matrix ${\bf D^{T, m}}$, thus obtaining the length of the shortest path connecting every pair of nodes in the network. Note that our method does not alter the structure of shortest paths in the network; therefore ${\bf D^{T,m}}$ constitutes the metric closure matrix of the final synthetic network.

On top of the metric backbone, we add $E_{sm}$ edges to tune its relative size $\tau^m$ compared to the total size of the constructed network. If $E=E_m +E_{sm}$ denotes the total number of edges in the constructed network, the relative size fulfills $E_m = \tau^m E$. Therefore, to fix a specific value $\tau^m$, the number of semi-metric edges to be added is:
\begin{equation}
E_{sm} = \frac{1-\tau^m}{\tau^m} E_m\ .
\label{eq:semi_metric}
\end{equation}
These $E_{sm}$ edges are chosen randomly within the set of edges not present in the metric backbone {\bf B}. \rev{We choose them randomly as this is an agnostic way of constructing simple synthetic networks  to unveil the limitations of the metric backbone.} Note that $\tau^m \in \left[\frac{2E_m}{N(N-1)},1\right]$, where the lower bound corresponds to including all missing semi-metric edges to obtain a fully-connected network. 
%
%rev{Since the metric backbone is the primary transmission subgraph \cite{Correia:2023contact}, which nodes the added semi-metric edges connect to will most likely have negligible effects compared to its semi-metric distortion.}
% Therefore, randomly adding semi-metric edges to create synthetic networks is a reasonable first approximation.

%For the sake of simplicity, we construct the network following the Bar\'ab\'asi-Albert model and draw the weights $\vec{w}$ of the generated connections from a target weight probability distribution $P(w)$. Note that other models can be used for this step, such as the configuration model, providing the existence of a single connected component is guaranteed \rev{Change explanation to match the metric backbone from the real networks}. The resulting network constitutes the metric backbone of the actual synthetic network we aim at generating.
Once we fix the relative size of the backbone, we move to tuning the distortion of the semi-metric edges in the network.
To do so, for each added edge, we sample its semi-metric distortion value $s_{ij}^m$ from the target distribution $P(s^m)$ and assign the individual distance $d_{ij}$ of the new link following Eq.~(\ref{eq:s}). Therefore,
\begin{equation}
    d_{ij}=s_{ij}^m d^{T, m}_{ij}\ ,
    \label{eq:dprime}
\end{equation}
which are eventually transformed into proximity values by using Eq.~(\ref{eq:d}), yielding:
\begin{equation}
    p_{ij} = \left(d_{ij} +1\right)^{-1}\ .
\label{eq:p}    
\end{equation}

\subsection{SI dynamics}
\label{sec:si_dynamics}

We focus on the SI compartmental model as a proxy for spreading dynamics due to its simplicity.
In the SI model, there are only two states (compartments) available for each individual: Susceptible and Infected. A susceptible host $i$ contracts a virus and becomes infected when interacting with one infectious counterpart $j$ at a rate $\beta p_{ji}$. Once nodes are infected, they remain in this state over the entire dynamics. \rev{Following those update rules, we perform agent-based simulations of the model.}
Therefore, the SI dynamics is entirely characterized by the distribution of times at which each agent $i$ becomes infected, $t_i$. Note that, as shown in~\cite{Correia:2023contact}, $\beta$ is not a relevant parameter to assess the metric backbone performance, as it just represents a global redefinition of the timescale of the spreading process. Throughout the manuscript, we fix $\beta=0.5$ and perform all the simulations considering a single individual initially infected, constituting the seed of the infection. Generally, we characterize the outbreak by the time at which half of the population is infected, denoted by $t_{1/2}$. This value is obtained for $50$ different initial seeds, to smooth out possible biases introduced by the origin of the outbreak in our analysis, and by averaging the results of $200$ realizations for each seed.

\subsection{Effective resistance}
\label{sec:effective_resistance}

One of the sparsification methods with which we compare the proposed \rev{SMDS} is the removal of connections following the effective resistance edge ranking~\cite{mercier2022effective}.
The effective resistance between two nodes $i$ and $j$ denoted by $R^e_{ij}$ captures their global exchange of information through all the different paths connecting them in the network. Mathematically, one computes the effective resistance by:
\begin{equation}
R^e_{ij}= (e_i-e_j)^T L^\dagger (e_i - e_j)\ ,
\end{equation}
where $L^\dagger$ represents the Moore-Penrose inverse of the Laplacian matrix of the network and $\vec{e}$ the elements of the canonical basis. The effective resistance has proven to remove connections while preserving spreading SIR dynamics. Specifically, one can define the probability of keeping the edge connecting nodes $i$ and $j$, $p^{R}_{ij}$, as~\cite{mercier2022effective}
\begin{equation}
p^{R}_{ij}\propto p_{ij}R^e_{ij}\ .
\label{eq:r}
\end{equation}
Thresholding the network according to the former probabilities prevents from isolating nodes, as $p^{R}_{ij} = 1$ when the edge $(i,j)$ represents the single path connecting both nodes. As more paths becomes available, the former value becomes smaller, thus penalizing redundancy of information flow among nodes. To preserve SI dynamics, we must then remove those edges with lowest effective resistance values, as less relevant transmission pathways are hampered following their removal. Note that the ordering for edge removal according to their effective resistance values is computed only once considering the original network as a reference. 

\subsection{Empirical Networks}
\label{sec:datasets} 

This study also uses 16 networks that were obtained from experimental data from social contacts (7), transportation (7), and nervous systems (2).
Their properties, including the distribution of proximity values, the relative size of the metric backbone, and the semi-metric distortion distribution, are available in Supplementary Figure S1 and Table \rev{S1}.
%
%The semi-metric distortion sparsification of these networks is possible due to their connectivity patterns and heterogeneity in edge weights, which we observe in Figure S1 in the Supplementary text.

All social contact networks considered were previously studied in \cite{Correia:2023contact}, and are available in the \href{http://www.sociopatterns.org/}{SocioPatterns Database} from their original studies \cite{toth2015role, stehle2011high, salathe2010high, vanhems2013estimating, mastrandrea2015contact, genois2018can}.
%, and made available by the authors in the \href{https://github.com/rionbr/socialbackbone}{Social Backbone GitHub} page.
%
Proximity edge strength values  are obtained via proximity sensors that measure the time spent by individuals in the vicinity of each other. Specifically, we tally the number of 20 minute intervals when each pair of individuals $i$ and $j$ were in direct contact with each other: $r_{ij}$. From this measurement, we compute the proximity matrix {\bf P}  defined in \S \ref{sec:backbone_method} according to the Jaccard measure \cite{Jaccard:1901} for each entry: 
\begin{equation}
    p_{ij} = \frac{r_{ij}}{r_{ii} + r_{jj} - r_{ij}}\ ,
    \label{eq:Prox_contact}
\end{equation}
\noindent where $r_{ii}$ tallies the total number of 20 minute intervals individual $i$ was present in the experiment. The isomorphic distance matrix {\bf D} is obtained via the equation~\ref{eq:d}.

Six transportation networks are similarly built from mobility data from  core-based statistical areas (CSBA) in the state of New York in the United States (US).
%, each of which comprise one network, and seat availability data in flights within the continental US (airports network).
%
%In the networks using CSBA data, the 
%nodes correspond to ZIP codes and edge weights are relative to the total number of trips between the nodes based on census surveys carried out in 2017. 
Proximity and distance matrices are built in the same way as the social contact networks via eqs. \ref{eq:Prox_contact} and ~\ref{eq:d}, but where $r_{ij}$ tallies the  number of trips between ZIP codes $i$ and $j$, which are cast as nodes.
A seventh transportation network is included to characterize  air traffic between 
the 500 busiest commercial airports in the US during 2002.
It is built in the same manner, with $r_{ij}$ tallying the number of available airplane seats between airports $i$ and $j$. 
%,  Its nodes are connected according to the amount of available seats between them normalized by the Jaccard measure \cite{Jaccard:1901} resulting in the proximity matrix {\bf P}.

Two nervous system networks are also studied: a \textit{human connectome }network mapping the strength of connections across 66 brain regions of interest \cite{Hagmann:2008}, and a the neural network of the \emph{Caenorhabditis elegans} worm (c-elegans) \cite{Watts:1998}. 
%
%Their backbone sparsification has been studied previously \cite{simas2021distance} also showing a log-normally distributed semi-metric distortion values.
%
Proximity and distance matrices are built in the same way as the social contact networks via eqs. \ref{eq:Prox_contact} and ~\ref{eq:d}, but where $r_{ij}$ tallies the 
%
%The former maps 66 regions of interest as nodes of a network with proximity weights {\bf P} between them given by the 
number of streamlines, identified via diffusion spectrum imaging, per region volume between brain regions of interest $i$ and $j$, using data from the human connectome extracted by \cite{Hagmann:2008}, 
and the number of gap junctions between pairs of neurons $i$ and $j$  in the c-elegans nervous system network \cite{Watts:1998}.
%, nodes in this network, to compute the proximity matrix {\bf P} according to the Jaccard measure.

\section{Acknowledgements}

We thank Rion Brattig Correia and the members of the CASCI lab for useful discussions about this project. This work was funded by NIH National Library of Medicine Program grant 01LM011945-01 to LMR and the Fundação para a Ciência e a Tecnologia grant 2022.09122.PTDC (\url{https://doi.org/10.54499/2022.09122.PTDC}) to LMR and FXC. DS-P was funded by funded by MCIN/AEI/10.13039/501100011033 and the European Union “NextGenerationEU”/PRTR” through grants JDC2022-048339-I and PID2021-128005NB- C21.
The funders had no role in any stage of the study design to the manuscript preparation, nor decision to publish.

 \newpage
\appendix
\renewcommand\thefigure{S\arabic{figure}}
\renewcommand\thetable{S\arabic{table}}
\setcounter{figure}{0}
\setcounter{table}{0}

\newpage
\renewcommand\thefigure{S\arabic{figure}}
\renewcommand\thetable{S\arabic{table}}
\setcounter{figure}{0}
\setcounter{table}{0}

\nolinenumbers

\section*{\Large{Supplementary text}}

\section*{Features of the empirical networks analyzed in the study}

\begin{table}[h!]
\centering
\caption{Properties of the empirical networks studied. We report the number of nodes ($N$) and edges ($E$), the corresponding density ($\delta$) and the relative size of the backbone $\tau^m$. The last two columns show the log median $\mu$ and dispersion $\sigma$ parameter obtained when fitting the semi-metric distortion distribution to a log-normal distribution. Distortion distribution curve fits were done using the \textsc{powerlaw} Python package}
    \begin{tabular}{c|c|c|c|c|c|c|c}
   {} & Network & $N$ & $E$ & $\delta$ & $\tau^m$ (\%) & $\mu$ ($s_{ij}$) & $\sigma$ ($s_{ij}$) \\
     \hline \hline
    %conference & 113 & 2196 & 0.173515 & 0.140255 & 1.507951 & 0.906073 \\
    & US Middle School & 591 & 56867 & 0.16 & 6.19 & 2.48 & 1.38 \\
    & US Elementary School & 339 & 16546 & 0.14 & 6.82 & 2.12 & 1.31 \\
    & US High School & 788 & 118291 & 0.19 & 7.84 & 2.84 & 1.56 \\
    & French Primary School & 242 & 8317 & 0.14 & 9.50 & 1.80 & 1.41 \\
    & French High School & 327 & 5818 & 0.05 & 10.36 & 2.24 & 1.48 \\    
    & Workplace & 217 & 4274 & 0.09 & 17.43 & 0.61 & 1.23 \\
    \hfill \begin{rotate}{90} Social \end{rotate} & Hospital & 75 & 1139 & 0.21 & 19.05 & 1.10 & 1.55 \\
    \hline
    & Albany, NY & 127 & 4622 & 0.29 & 8.39 & 1.16 & 1.44 \\
    & Poughkeepsie, NY & 86 & 2004 & 0.27 & 8.78 & 1.38 & 1.42 \\
    & Rochester, NY & 122 & 4872 & 0.33 & 9.75 & 1.44 & 1.60 \\
    & Utica, NY & 73 & 1432 & 0.27 & 10.61 & 0.95 & 1.35 \\
    & Syracuse, NY & 103 & 3304 & 0.31 & 12.17 & 1.21 & 1.50 \\
    & Buffalo, NY & 90 & 3237 & 0.40 & 12.26 & 1.15 & 1.63 \\
    \hfill \begin{rotate}{90} Transportation \end{rotate} & Airports & 500 & 2980 & 0.01 & 37.15 & 0.32 & 1.96 \\
    \hline
    & Human Connectome & 66 & 1148 & 0.27 & 9.23 & 1.45 & 1.10 \\
    \hfill \begin{rotate}{90} Brain \end{rotate} & c-elegans & 297 & 2148 & 0.02 & 46.97 & 1.125 & 0.004 \\
    
    \hline
    \end{tabular}
\label{tab:nets_description}
\end{table}

\clearpage
\begin{figure}[t!]
    \centering
    \includegraphics[width=1.00\columnwidth]{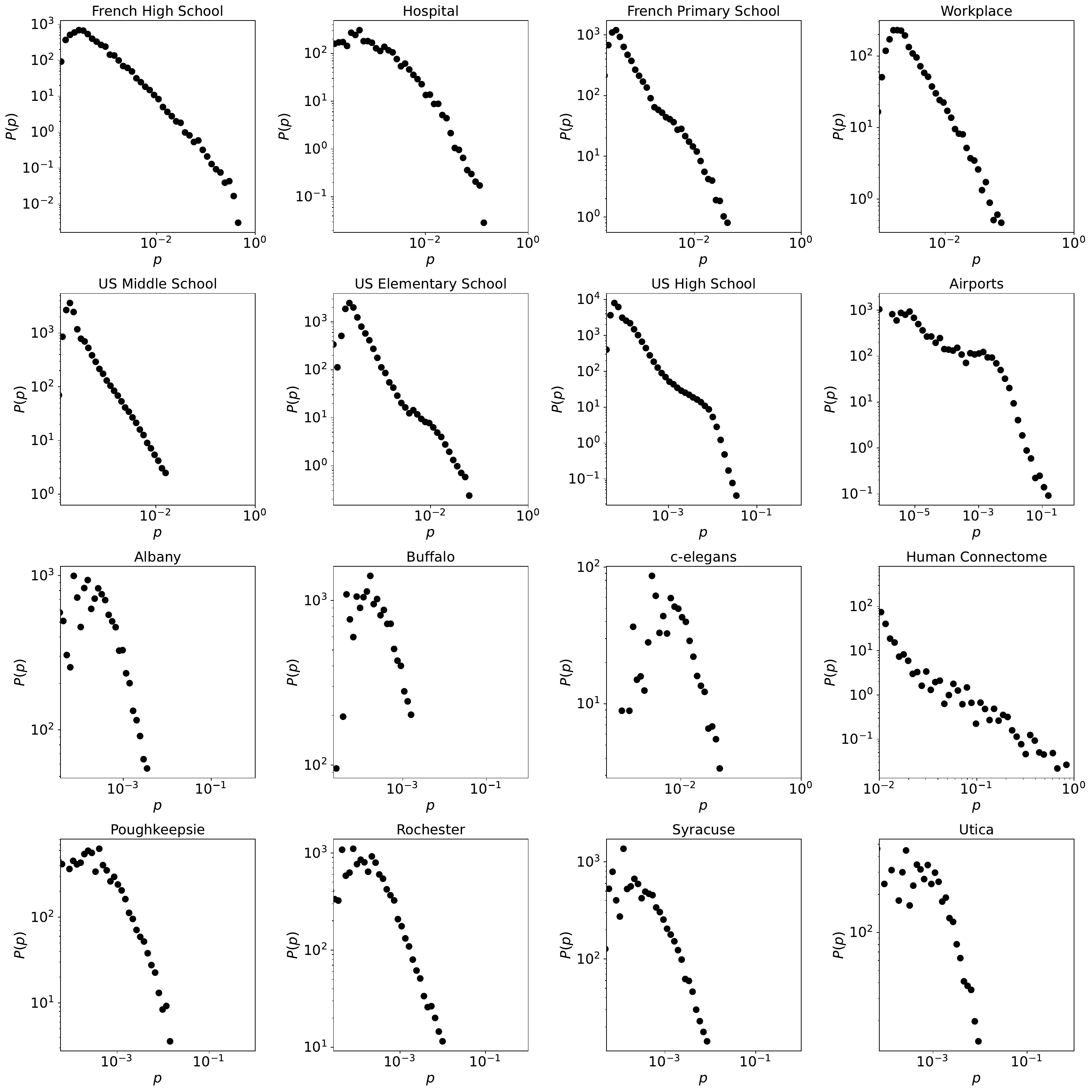}
    \caption{Distribution of proximity values $p$  for the edges in the metric backbone of the empirical networks described in Table~\ref{tab:nets_description}.}
    \label{fig:proximity_metric_edges}
\end{figure}

\clearpage

\section*{Impact of different sparsification methods on the metric backbone}
\begin{figure}[h!]
    \centering
    \includegraphics[width=0.8\columnwidth]{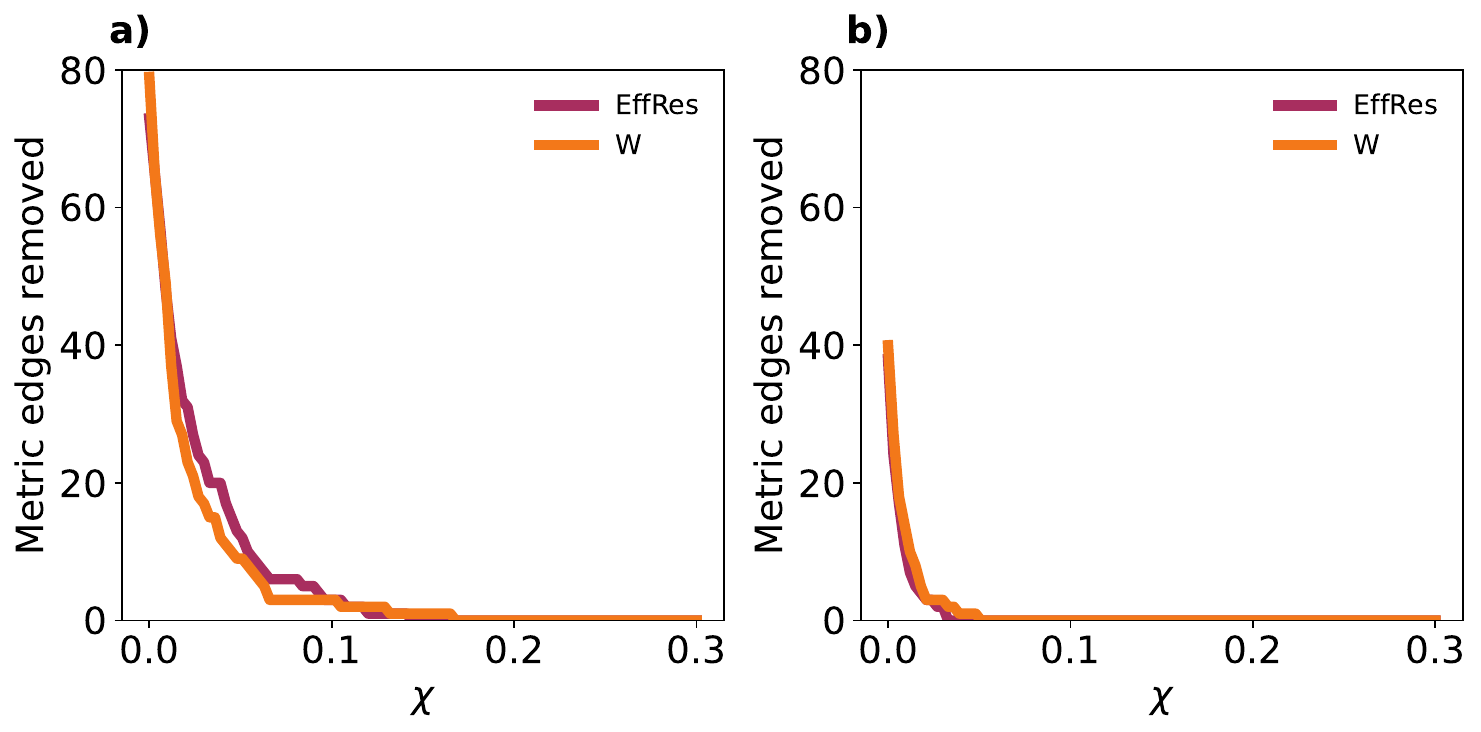}
    \caption{Number of metric edges removed for synthetic networks relative to the size of the sparsified network governed by $\chi$. $\chi$ spans from $\chi=1$, corresponding to not removing any edges, to $\chi=0$ where $E^{sm}$ edges are removed. In the synthetic networks the backbone is fixed as well as the dispersion parameter, $\sigma=1$, of the log-normally distributed semi-metric distortion values. Regarding the log median of the distribution the values used are (a) $\mu=1$ and (b) $\mu=2$.}
    \label{fig:target_metric_edges}
\end{figure}
\clearpage

\section*{Semi-metric distortion sparsification and SI dynamics}
\begin{figure}[h!]
    \centering
    \includegraphics[width=1.00\columnwidth]{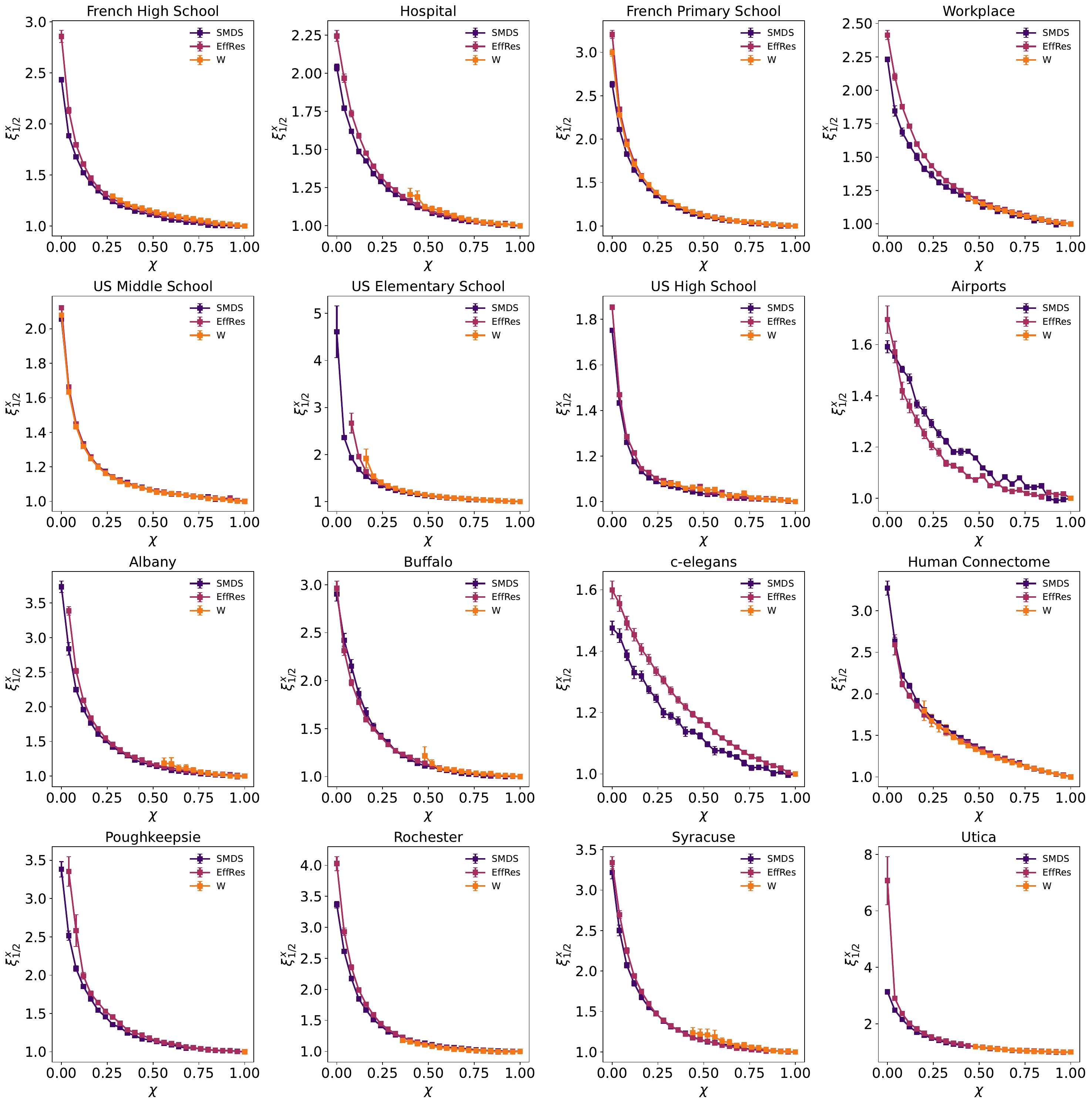}
    \caption{Ratio between the time for the disease to reach half the population in the sparsified network and in the empirical networks $\xi^{x}_{1/2}$  for each sparsification method $x$ (color code) as a function of the parameter $\chi$, governing the edges removed from the network. $\chi$ spans from $\chi=1$, corresponding to not removing any edges, to $\chi=0$ where $E_{sm}$ edges are removed. Three different sparsification methods are compared: targeting edges with highest semi-metric distortion values (SMDS), lowest effective resistance values (EffRes) or lowest proximity values (W).
    The details of the simulations to obtain the ratios are the ones described in the main text.
    EffRes and W curves are interrupted when edges removal breaks the largest connected component of the network.}
    \label{fig:all_networks}
\end{figure}

\begin{figure}[t!]
\includegraphics[width=.8\columnwidth]{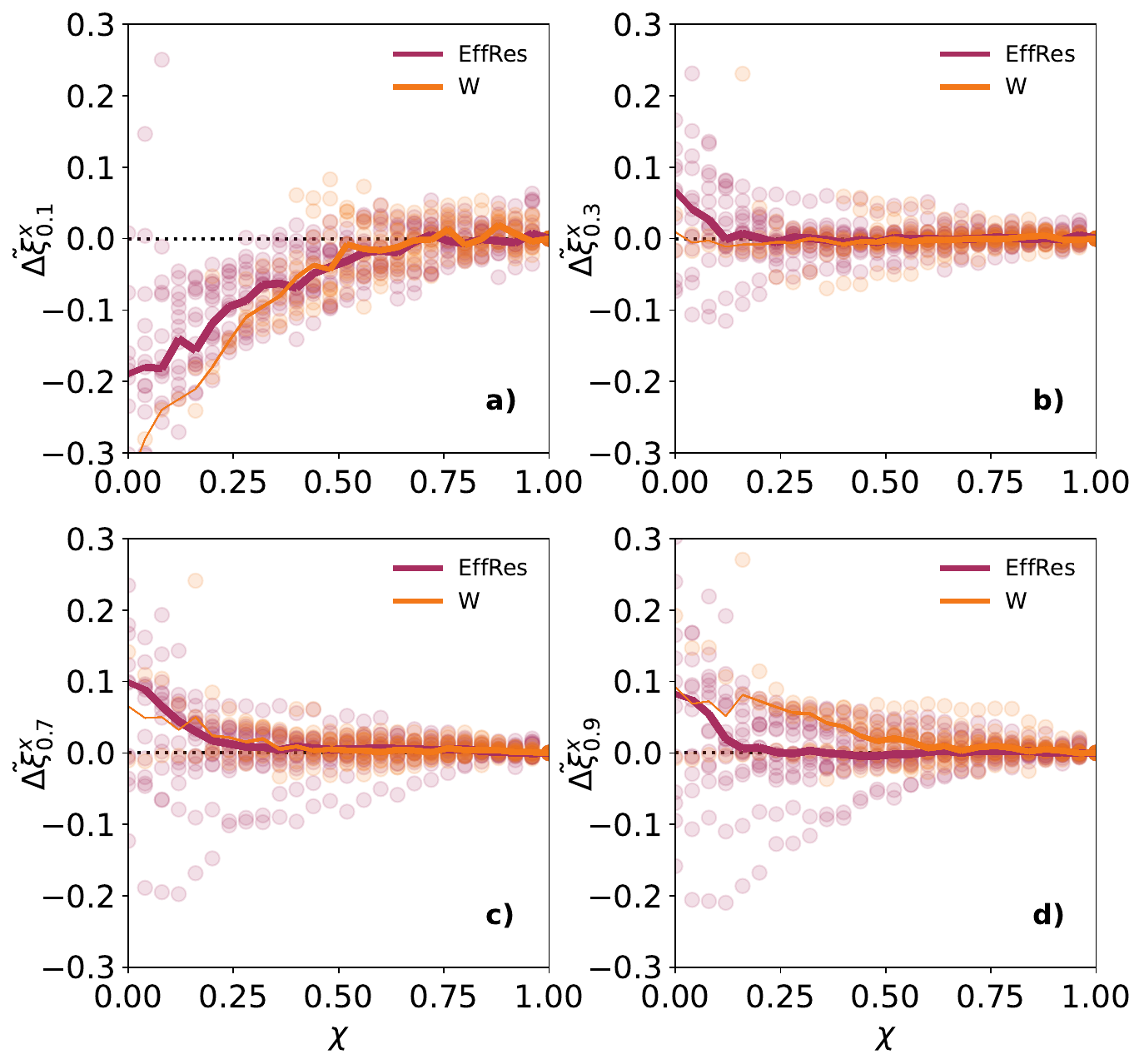}
\caption{Distribution of the relative difference $\tilde{\Delta\xi}^x_y$ between the time that the outbreak takes to reach a given fraction $y$ of the population, comparing each sparsification method $x$ (color code) with SMDS (black dashed line), as a function of the size of the sparsified network $\chi$. $\chi$=1 preserves all the connections in the network whereas $\chi=0$ corresponds to subgraphs whose number of connections matches the size of the metric backbone. Positive $\tilde{\Delta\xi}^x_{y}$ values reveal a better performance of SMDS with respect to the other two methods. Dots represent each of the 16 networks included in the dataset under study whereas solid lines show the median $\tilde{\Delta\xi}^x_{y}$ value observed across the set of empirical networks. Line thickness is proportional to the fraction of networks remaining connected after sparsification (shown in Fig. 4c in the main text). The fraction of the population $y$ used in each panel are: $10\%$ (panel a), $30\%$ (panel b), $70\%$ (panel c) and $90\%$ (panel d).}
\label{fig:Supp_diff_times}
\end{figure}
\clearpage

\section*{Impact of sparsification on network connectivity}
\begin{figure}[h!]
    \centering
    \includegraphics[width=0.8\columnwidth]{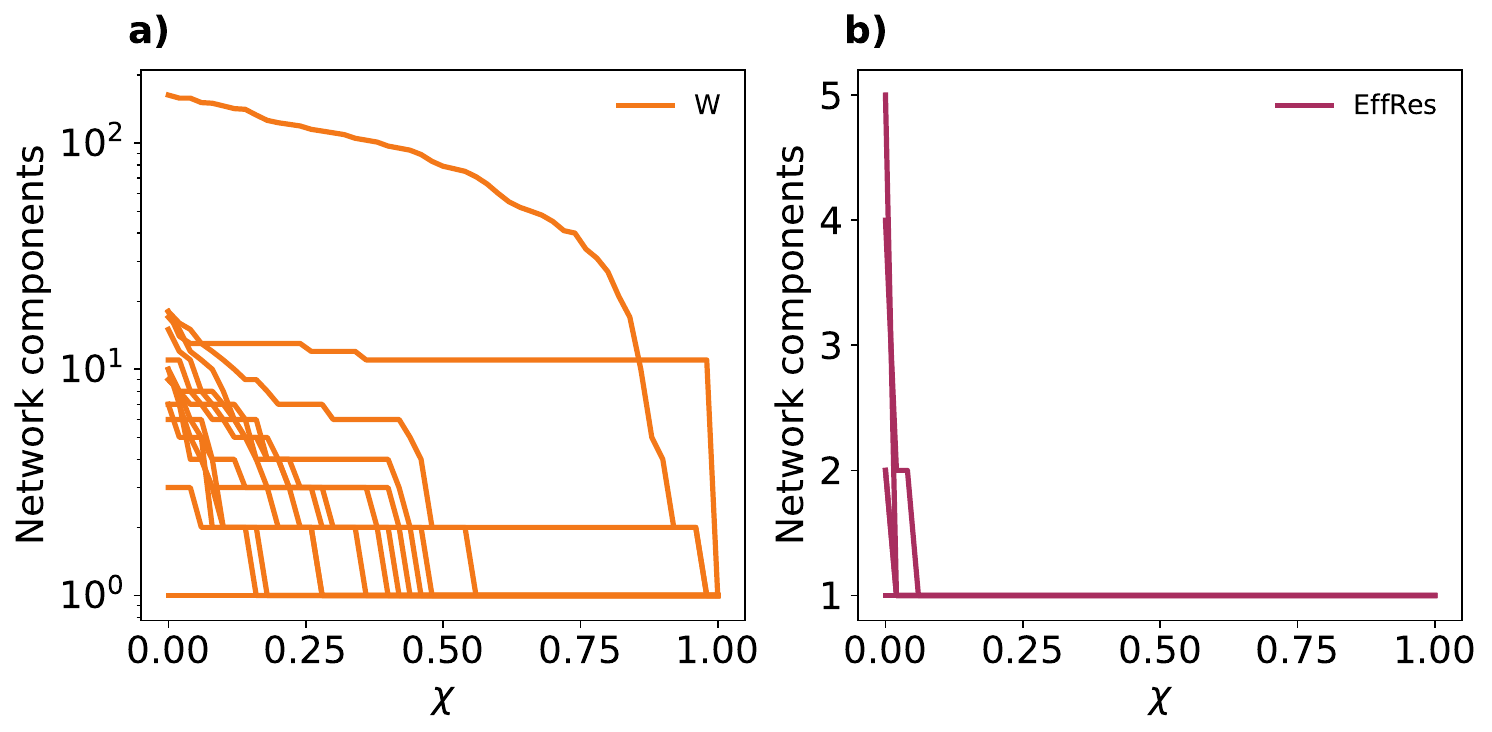}
    \caption{Number of connected components for the empirical networks as a function of the size of the sparsified network $\chi$. $\chi$=1 preserves all the connections in the network whereas $\chi=0$ corresponds to subgraphs whose number of connections matches the size of the metric backbone.}
    \label{fig:components}
\end{figure}

\clearpage
\section*{Semi-metric Distortion Sparsification in other compartmental models}
The analysis presented in the main text to address whether semi-metric distortion sparsification (SMDS) allows sparsifying networks while fairly preserving spreading dynamics focuses on the Susceptible-Infected (SI) dynamics. However, such model assumes individuals remain infectious throughout the entire course of an epidemic outbreak once they have contracted the disease. Here we extend our analysis to more realistic compartmental models accounting for the recovery of infected agents. In particular, we focus on two of the most paradigmatic models used in mathematical epidemiology, the Susceptible-Infected-Susceptible (SIS) and the Susceptible-Infected-Recovered (SIR) models.

\subsection*{SIS dynamics}
The SIS model assumes that each agent can be either in the Susceptible state or in the Infected state. Moreover, the SIS model assumes that a susceptible agent can contract the disease upon contact with an infected agent at a rate $\beta$ whereas infected agents overcome the disease at a rate $\mu$, becoming again susceptible to contract the disease. Therefore, the SIS model applies to those diseases for which contracting the pathogen does not confer immunity to the host for future infections. 

In the main text, we use agent-based simulations to generate epidemic outbreaks, as the order parameter of the SI model should be computed during its transient dynamics as the time to reach a given percentage of the population. In contrast, the order parameter of the SIS dynamics is the epidemic prevalence $\rho^*$, defined as the fraction of the population in the infected compartment  once the dynamics has reached its steady state. Therefore, to assess the performance of sparsification methods, we can analyze how they alter the steady state of the SIS model through the numerical integration of ODEs, considerably reducing the computation time with respect to agent-based simulations.

Among the different theoretical models proposed to simulate the SIS dynamics, here we resort to the Quenched Mean-Field (QMF) equations. Under this approach, the dynamics is characterized by the probabilities that each individual $i$ is in the infected state at time $t$, hereinafter denoted by $\rho_i (t)$. Recalling that the proximity between agents $i$ and $j$ is given by the element $p_{ij}$ of the matrix ${\bf P}$, the time evolution of those variables reads:
\begin{equation}
    \dot{\rho}_{i}=-\mu \rho_i + (1-\rho_i)\beta\sum\limits_j p_{ij}\rho_j\ .
\end{equation}

Consequently, the order parameter for the SIS dynamics is defined as $\rho^* = \frac{1}{N}\sum\limits_i \rho_i^*$, where $\rho_i^*$ refers to the probability that the agent $i$ is in the infected compartment at the steady state of the dynamics.

Figure S6 shows how $\rho^{*,x}$ varies as a function of the size of the sparsified network for the empirical networks used in this study according to each sparsification method $x$ discussed in the main text. In all panels, the epidemiological parameters are set to $\beta=0.3$ and $\mu=0.1$ ensuring the existence of an endemic state for the actual networks, i.e. when $\chi=1$. For all networks and sparsification methods, we observe that the removal of connections reduces the prevalence of the disease in the endemic state, leading in some cases to its eventual extinction for small sizes of the sparsified network. This is an expected result, as removing connections interrupts potential transmission pathways required to maintain widespread epidemic states in networks. 

Comparing across sparsification methods, we notice that weights thresholding, removing the weakest connections, slightly outperforms the other two methods in retrieving the epidemic state when removing few connections. Note that this contrasts with the results obtained for the SI dynamics in which weights thresholding was the one providing the worst retrieval of the times to reach half of the population. Therefore, while preserving global spreading pathways is more relevant to understand the diffusion of epidemics in the SI model, keeping strong connections is more crucial to maintaining locally stable epidemic states. As more connections are removed, weights thresholding loses suitability, as it breaks down most of the network into different subcomponents. In this scenario, both effective resistance thresholding and SMDS yield comparable results in retrieving the global epidemic state of the system.

Microscopically, we can also assess the performance of the different sparsification methods by computing $\rho_i^* (\chi)$, thus measuring how the size of the sparsified network, governed by $\chi$, alters the individual probabilities of being infected in the steady state. To quantify such performance, we measure the Spearman correlation coefficient between the spatial distribution of individual prevalence of the original network, i.e. $\left\lbrace \rho_i^* (\chi=1)\right\rbrace$ with that observed in the sparsified configuration according to the sparsification method $x$, i.e.$\left\lbrace \rho_i^{*,x} (\chi)\right\rbrace$. Figure S7 shows how effective resistance thresholding and SMDS preserve the spatial ranking of epidemic prevalence until $\chi\simeq 0.20$ and that SMDS generally outperforms effective resistance thresholding in retrieving the microscopic features of the epidemic state. The performance drop of both methods for $\chi \simeq 0$ is caused by the change in the nature of the epidemic state, as the subsequent removal of connections turns widespread epidemics into localized epidemic states in the most vulnerable areas of the network. 

\begin{figure}[h!]
    \centering
    \includegraphics[width=1\columnwidth]{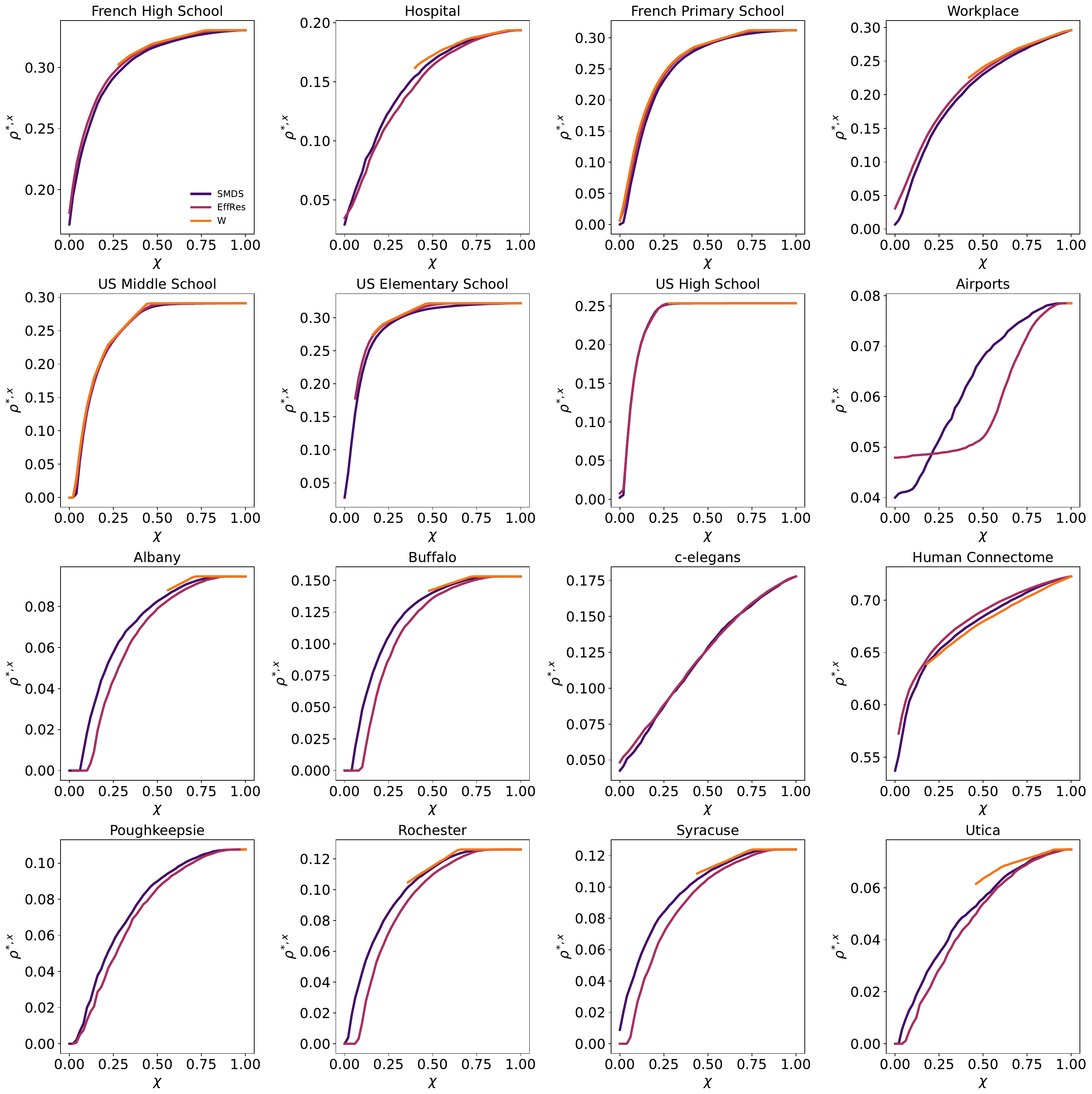}
    \caption{Fraction of individuals in the infected state in the steady state $\rho^{*,x}$ of a SIS dynamics as a function of the size of the sparsified network, governed by $\chi$, and the sparsification method $x$ (color code). Each panel shows the results for a real network of those detailed in Table~\ref{tab:nets_description}. In all panels, the infectivity rate has been set to $\beta=0.3$ and the recovery rate to $\mu=0.1$. Error bars are not included as the endemic equilibrium of the system of ODE is unique for each $\chi$ value and combination of epidemiological parameters.}
    \label{fig:SISorderparameter}
\end{figure}

\begin{figure}[h!]
    \centering
    \includegraphics[width=1\columnwidth]{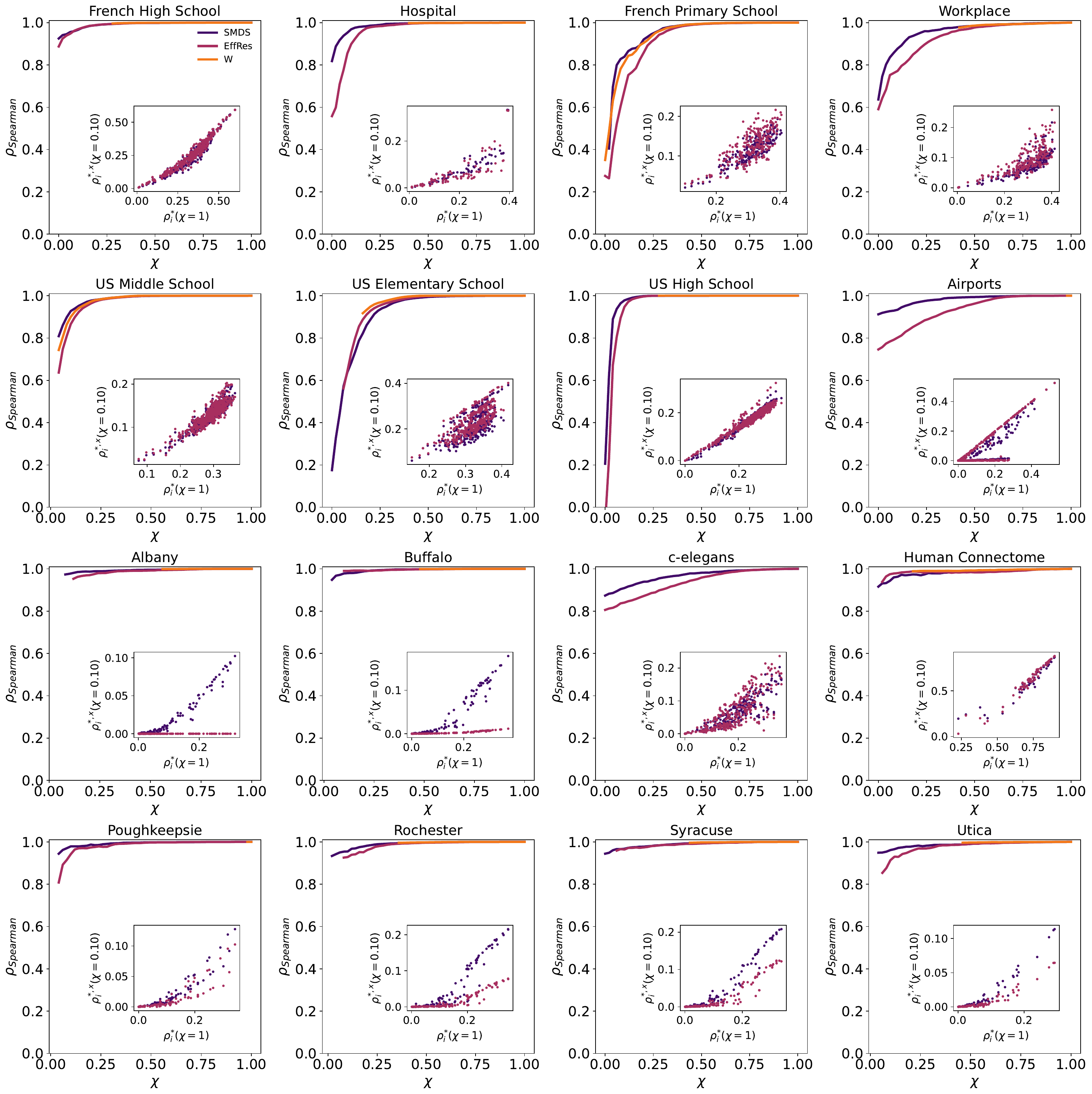}
    \caption{Spearman correlation coefficient between the distribution of the probabilities that each node $i$ is infected at the steady state of a SIR dynamics in the original network, $\left\lbrace r_i^{*} (\chi=1)\right\rbrace$, and that of the sparsified network ,$\left\lbrace r_i^{*,x} (\chi)\right\rbrace$, as a function of the size of the sparsified network $\chi$ and the sparsification method $x$ (color code). Fraction of individuals in the infected state in the steady state $\rho^{*,x}$ of a SIS dynamics as a function of the size of the sparsified network, governed by $\chi$, and the sparsification method $x$ (color code). Inset: Scatter plot relating the probability of infection in the sparsified network with sparsification method $x$ (color code) and $\chi=0.1$, $\left\lbrace \rho_i^{*,x} (\chi=0.1)\right\rbrace$,  as a function of the probability of infection of the node in the steady state in the original network, $\left\lbrace \rho_i^{*} (\chi=1)\right\rbrace$.  Note that weights thresholding is not represented in the scatter plots as most of the networks do not preserve the largest connected component for ${\chi=0.1}$ (Fig. S5). Each panel shows the results for a real network of those detailed in Table~\ref{tab:nets_description}. In all panels, the infectivity rate has been set to $\beta=0.3$ and the recovery rate to $\mu=0.1$.}
    \label{fig:SISmicroscopic}
\end{figure}

\clearpage
\subsection*{SIR dynamics}

The SIR model assumes that each agent can be in three states: susceptible to contract the disease (S), infected (I) or recovered (R), gathering those who were infected at some point of the dynamics and are immune against future infections. In mathematical terms, the SIR model assumes that infected individuals recover at a rate $\mu$, entering the R compartment whereas susceptible individuals contract the disease at a rate $\beta$ after contact with their infectious neighbors. Defining $r_i (t)$ as the probability that an agent $i$ occupies the compartment R at a time $t$, the QMF equations for the SIR dynamics read:
\begin{eqnarray}
\dot{\rho}_i &=& -\mu \rho_i + (1-\rho_i-r_i)\beta \sum\limits_j p_{ij}\rho_j\ , \\
\dot{r}_i &=& \mu \rho_i\ .
\end{eqnarray}

From the equations of the SIR model, it becomes clear that its steady state is characterized by the absence of infected individuals. To characterize the extent of an epidemic outbreak, the order parameter of the SIR dynamics is the attack rate $r^*$, defined as the fraction of the population in the recovered compartment at the end of the outbreak. Figure S8 represents how the different sparsification methods preserve this quantity, obtaining results similar to those reported for the SIS dynamics. 

Microscopically, the SIR dynamics can be characterized by studying the probability that each node $i$ has been infected at some point in the epidemic outbreak. This information is reflected in the probability $r_i^*$ that this node $i$ occupies compartment R at the end of the outbreak. The results, shown in Figure S9,are consistent with those reported for the SIS dynamics, showing a better performance of the SMDS compared to the other methods discussed here in ranking the nodes according to their vulnerability. 

\begin{figure}[t!]
    \centering
    \includegraphics[width=1\columnwidth]{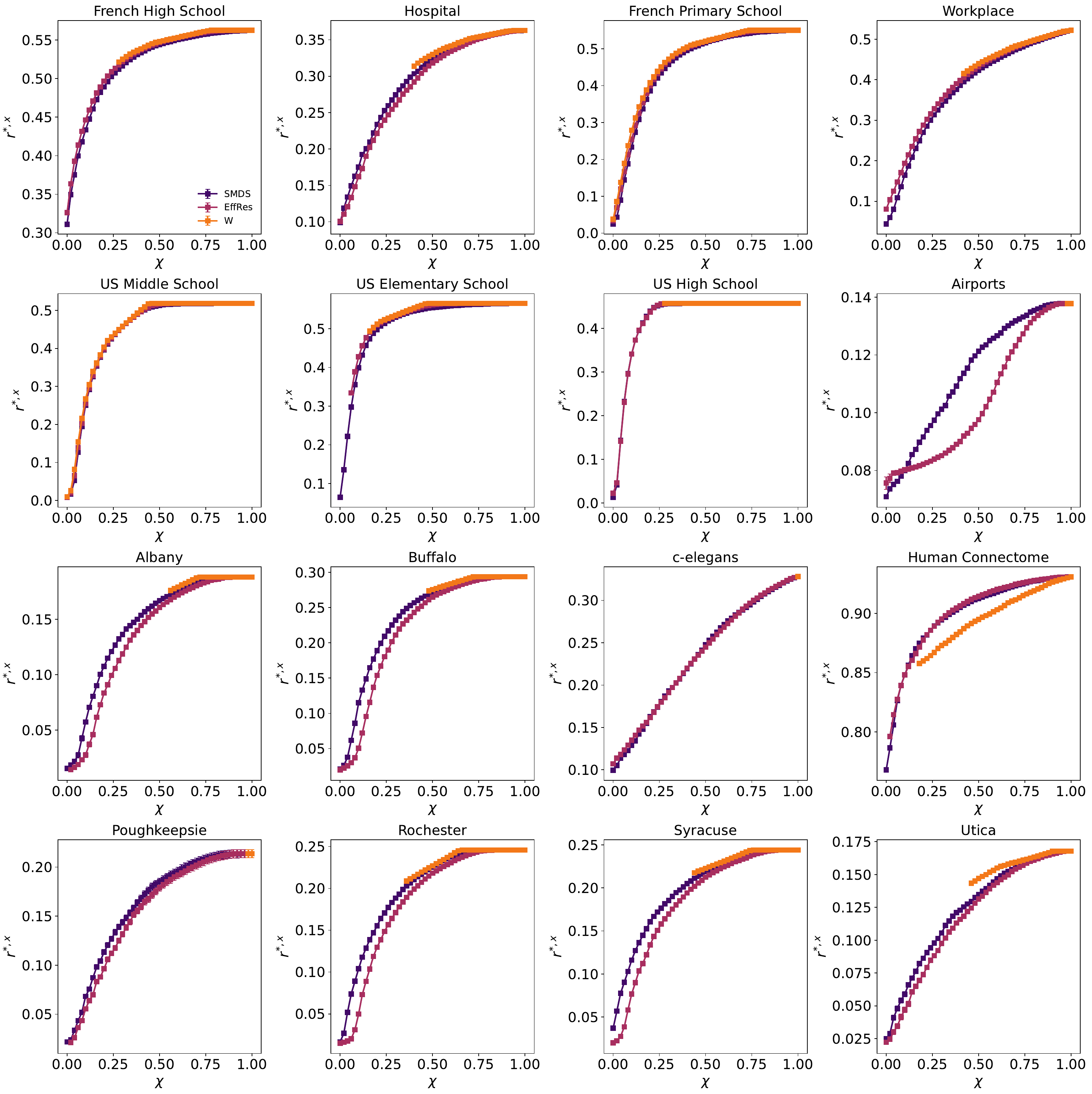}
    \caption{Fraction of individuals in the recovered state in the steady state $r^{*,x}$ of a SIR dynamics as a function of the size of the sparsified network, governed by $\chi$, and the sparsification method $x$ (color code). Each panel shows the results for a real network of those detailed in Table~\ref{tab:nets_description}. In all panels, dots represent the order parameter averaged across $50$ realizations, each one starting by a single randomly chosen individual, and the error bars show its standard deviation. Regarding epidemiological parameters, the infectivity rate has been set to $\beta=0.3$ and the recovery rate to $\mu=0.1$. }
    \label{fig:SIRorderparameter}
\end{figure}

\begin{figure}[t!]
    \centering
    \includegraphics[width=1\columnwidth]{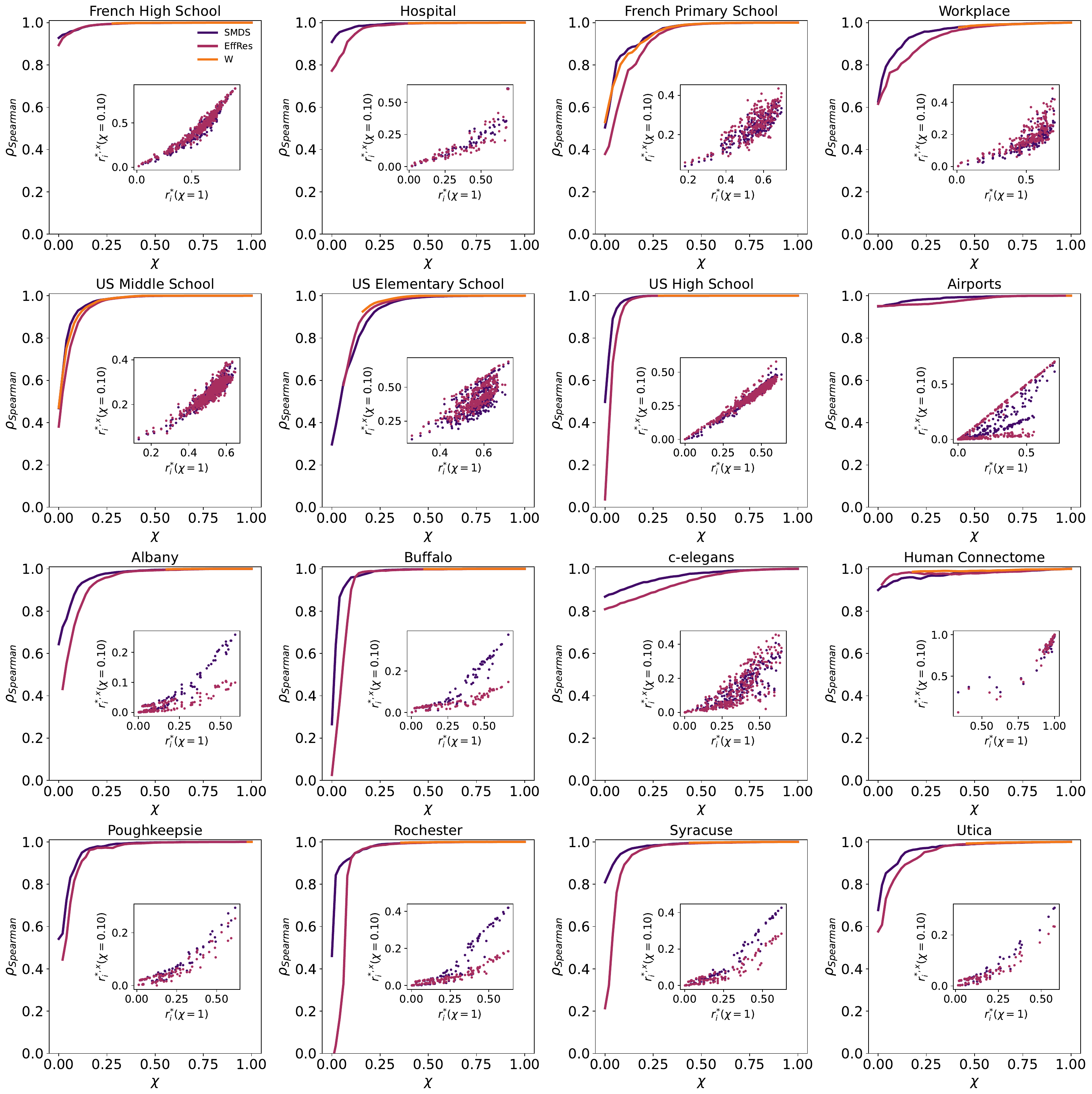}
    \caption{Spearman correlation coefficient between the distribution of the probabilities that each node $i$ is recovered at the steady state of a SIR dynamics in the original network, $\left\lbrace r_i^{*} (\chi=1)\right\rbrace$, and that of the sparsified network ,$\left\lbrace r_i^{*,x} (\chi)\right\rbrace$, as a function of the size of the sparsified network $\chi$ and the sparsification method $x$ (color code). Inset: Scatter plot relating the average probability of infection for node $i$ in the sparsified network with sparsification method $x$ (color code) and $\chi=0.1$, $\left\lbrace \rho_i^{*,x} (\chi=0.1)\right\rbrace$,  as a function of that probability observed the original network, $\left\lbrace \rho_i^{*} (\chi=1)\right\rbrace$.  Each panel shows the results for a real network of those detailed in Table~\ref{tab:nets_description}. Note that weights thresholding is not represented in the scatter plots as most of the networks do not preserve the largest connected component for ${\chi=0.1}$ (Fig. S5). In all panels, the infectivity rate has been set to $\beta=0.3$ and the recovery rate to $\mu=0.1$.}
    \label{fig:SIRmicroscopic}
\end{figure}

%\clearpage
%\bibliography{references}

\end{document}